\documentclass{aa}  

\usepackage{graphicx}
\usepackage{txfonts}
\usepackage{hyperref}
\usepackage{natbib}
\usepackage{makecell}
\usepackage{booktabs}
\usepackage{amsmath}
\usepackage{mathtools}

\newcommand{\mesa}{\texttt{MESA}\xspace}
\newcommand{\cpm}{\texttt{convective\_premixing}\xspace}
\newcommand{\gm}{\texttt{gentle\_mixing}\xspace}

\usepackage[print-unity-mantissa = false, range-units = single]{siunitx}
\usepackage{esdiff} % for easy derivatives

\newcommand{\pder}[3]{\left.\frac{\partial#1}{\partial#2}\right|_{\mathrlap{#3}}}
\newcommand{\altpder}[3]{\left.\partial#1 / \partial#2\right|_{\mathrlap{#3}}}

\DeclareSIUnit{\RJ}{R_J}
\DeclareSIUnit{\MJ}{M_J}
\DeclareSIUnit{\ME}{M_E}
\DeclareSIUnit{\kb}{k_B}
\DeclareSIUnit{\yr}{yr}
\DeclareSIUnit{\Gyr}{\giga \yr}
\DeclareSIUnit{\bary}{m_u}
\DeclareSIUnit{\kbperbary}{\kb \per \bary}

\graphicspath{{./images/}
\bibpunct{(}{)}{;}{a}{}{,}}
\hypersetup{
    colorlinks=true,
    linkcolor=blue,
    citecolor=blue,
    filecolor=magenta,      
    urlcolor=blue,
}

\begin{document}

\title{Giant planet evolution with \mesa}

\author{
    Ravit Helled\inst{1}
    \and
    Simon Müller\inst{1}
    \and
    Henrik Knierim\inst{1}
}
\authorrunning{Helled, Müller \& Knierim}

\institute{
    Department of Astrophysics, University of Zürich, \\
    Winterthurerstrasse 190, 8057 Zürich, Switzerland \\
    \email{ravit.helled@uzh.ch}
}

\date{Received; accepted}

\abstract{
The evolution of gaseous planets is a complex process influenced by various physical parameters and processes. In this study, we present critical modifications to the Modules for Experiments in Stellar Astrophysics (\mesa) code to enhance its applicability to giant planet modeling. We introduce an equation of state (EoS) specifically tailored for materials at planetary conditions. The equation of state considers the thermodynamic properties of hydrogen-helium mixtures and heavy elements, improving the accuracy of internal structure calculations. We also present modifications to the radiative opacity to allow the modeling of grains, clouds and opacity windows. Furthermore, we refine the treatment of convective mixing processes in \mesa to better replicate convective mixing with the presence of composition gradients. Finally, we add a treatment for helium rain and settling. These modifications aim to enhance the predictive capabilities of \mesa for giant planet evolution and are publicly available. We hope that these improvements will lead to a deeper understanding of giant planet evolution in the Solar System and beyond.
}


\maketitle

\section{Introduction}\label{sec:introduction}

The evolution of giant planets plays a crucial role in our understanding of planetary formation and internal structure \citep{2001RvMP...73..719B,2002ARA&A..40..103H,2008A&A...482..315B,2014prpl.conf..643H}. As we continue to discover exoplanets and deepen our insight into our Solar System's giants, it becomes increasingly important to advance theoretical models and numerical simulations in order to interpret the rich data.  
The evolution of giant planets is typically modeled by solving the partial differential equations of mass, momentum, and energy conservation, as well as the transport of chemical elements and energy \citep[e.g.,][]{2013sse..book.....K}.    
Several stellar models have been adapted to model planetary evolution such as \texttt{CEPAM} \citep[][]{guillot_cepam_1995} and \texttt{Etoile} \citep[][]{Kovetz2009, vazan_effect_2013}. Recently, a new planetary evolution code named \texttt{APPLE} that includes various relevant processes and parameters was presented \citep{2024ApJ...971..104S}. Unfortunately, these mentioned models are currently not public and unavailable to the scientific community. 

On the other hand, the Modules for Experiments in Stellar Astrophysics code (\mesa; \citet{2011ApJS..192....3P,2013ApJS..208....4P,2015ApJS..220...15P,2019ApJS..243...10P,2023ApJS..265...15J}) is open-source and is widely utilized for stellar evolution simulations and simple planet models. However, enhancing its capability and relevant for giant planet modeling requires certain modifications. This paper presents several significant modifications to the \mesa code (version 24.03.1). This allows the community to model the evolution of giant planets, accounting for the relevant physical processes and parameters. Our modifications include an equation of state (EoS) tailored for planetary materials, modifications of the radiative opacity, an improved treatment of convective mixing processes, and several methods of modeling helium rain or sedimentation of chemical species in general. By implementing these changes, we aim to produce more reliable models of giant planet evolution, thereby providing valuable insights into their formation and interior.

This paper does not aim at presenting new results; instead, it provides the technical details of the modifications to the \mesa code that make it a state-of-the-art planetary evolution model. The goal of this paper is to make this model accessible to the community, facilitating further research into the evolution and interiors of giant planets. 

Our paper is organized as follows. In Section \ref{sec:mesa_eos} we present the modifications to the EoS, while in Section \ref{sec:mesa_opacity} the opacity treatment. Sections \ref{sec:mixing} and \ref{sec:helium} describe the treatments of convective mixing and helium rain, respectively. 
A short summary is presented in Section \ref{sec:conclusions}. The code is open source and is publicly available as a GitHub repository\footnote{ \url{https://github.com/uzhplanets/mespa}}.

\section{A planetary equation of state for \mesa}\label{sec:mesa_eos}

A foundational aspect of modeling giant planets is the EoS, which describes how the properties of a material change under varying conditions. Since \mesa was primarily developed to model the evolution of stars, the calculations rely on equations of state constructed mainly for stellar matter, which can lead to inaccuracies and limitations when applied to planets. One major issue is that for planetary conditions, the EoS tables provided with \mesa are limited to heavy-element mass fractions of at most 4\%. This limits the range of models that can be calculated, since the deep interiors and envelopes of giant planets can be much more metal-rich. To address this limitation, we incorporated a more suitable EoS that reflects the unique conditions within giant planets, including hydrogen-helium mixtures and other complex components found in their atmospheres and interiors.

In this work, we present our most recent version of a planetary EoS developed to be used with \mesa. Earlier versions of the EoS were already used to investigate, for example, the origin of the dilute core in Jupiter \citep{2019Natur.572..355L, 2020A&A...638A.121M, 2025arXiv250323997M}, the planet's enriched atmosphere \citep{2024ApJ...967....7M}, mixing in giant planets \citep{Knierim2024, Knierim2025}, the characterization of giant exoplanets \citep[][]{2020ApJ...903..147M, muller_synthetic_2021, muller_towards_2023, 2025A&A...693L...4M}, and planet formation models \citep{valletta_giant_2020,2024A&A...685A..22M,2025arXiv250701212S}. 

Our EoS combines tables of single materials based on experimental data and theoretical calculations. Integrating this EoS into \mesa involves (i) constructing new tables and (ii) modifying the existing EoS subroutines to ensure appropriate handling of the new tables for a large range of different compositions.

\subsection{Construction}\label{sec:mesa_eos_construction}

We adapted the widely used linear-mixing approximation (Amagat's law) to construct an EoS for several components. The following relations can be readily derived by considering the Gibbs free energy of a mixture $G(p, T, \vec{N}) = \sum_i \mu_i N_i$ as a thermodynamic potential, where $p$ is the pressure, $T$ the temperature, and $\mu_i$ are the chemical potentials. The absolute numbers of each component $N_i$ are contained within the vector $\vec{N}$. For an arbitrary mixture, the total density $\rho$ of the mixture is given by  

\begin{equation}
    \rho^{-1}(p, T, \vec{X}) = \sum_i X_i \rho_i^{-1}(p, T) \, ,
    \label{eq:linear_mixing_density}
\end{equation}

\noindent
where $X_i$ and $\rho_i$ are the mass fraction and density of component $i$, the components of $\vec{X}$ are the individual mass fractions and $\sum_i X_i = 1$. It is further assumed that the mixture is in thermodynamic equilibrium, such that each component is at the same temperature and pressure. The total internal energy, $u$, and entropy, $s$, were calculated as the weighted sum of the individual quantities with an additional term for the entropy 

\begin{subequations}
    \begin{align}
        u(p, T, \vec{X}) &= \sum_i X_i u_i(p, T) \, \, \\
        s(p, T, \vec{X}) &= \sum_i X_i s_i(p, T)  + s_{\textrm{id}} \, ,
    \end{align}
    \label{eq:linear_mixing_energy_entropy}
\end{subequations}

\noindent
where $u_i$ and $s_i$ are the individual internal energies and entropies. Neglecting the free electrons, the ideal entropy of mixing  $s_{\textrm{id}}$ can be approximated as \citep[][]{2019ApJ...872...51C}: 

\begin{align}
     s_{\textrm{id}} &= k_b \left(N \ln N - \sum_i N_i \, \ln N_i \right) = -\frac{k_b}{\overline{A} m_H}\sum_i x_i \ln x_i \, ,
\end{align}

\noindent
where $N = \sum_i N_i$, $x_i = N_i / N$ are the number fractions, $k_b$ is the Boltzmann constant and $m_H$ the atomic hydrogen mass. The mean atomic weight is $\overline{A} = \sum_i x_i A_i$, where $A_i$ are the individual atomic mass numbers.

Following these rules, all necessary thermodynamic quantities to solve the evolution equations can be calculated, usually in more than one way. In Appendix \ref{sec:appendix_equation_of_state}, we describe the calculation of all quantities required by the EoS module of \mesa.

Due to its stellar-evolution heritage, \mesa is set up to handle a three-component EoS given the hydrogen, helium and heavy-element mass fractions $(X, Y, Z)$. This imposes a limitation that only one representative heavy-element component can be modeled in the interior. However, this representative does not need to be a pure substance, but can be any arbitrary mixture of several heavy elements. For our purposes, this was achieved by combining different individual heavy-element equations of state with the ideal mixing approximation. 

\mesa requires the logarithmic density and temperature as the thermodynamic basis, such that the EoS can be written as, for example, $p = p(\log \rho,\log T, X, Y, Z)$. The \mesa tables use the linear combination $\log Q = \log \rho \, [\rm{g/cc}] - 2 \log T \, [\rm{K]} + 12$ in place of the density as the first independent variable. This simplifies the interpolation, since most stellar and planetary interiors tend to be close to a diagonal in the $(\log \rho, \log T)$ space. Using $(\log \rho, \log T)$ as the independent variables complicates the application of the ideal mixing approximation, since the individual densities $\rho_i$ are unknown \textit{a priori}. Calculating the EoS at runtime would require a root-finding procedure for each EoS call, which is numerically unfeasible. Therefore, \mesa uses a set of pre-computed tables for different compositions, and then interpolates between these. Our EoS adheres to this strategy; however, since it spans heavy-element fractions from zero to one, it necessitates the computation of numerous tables.

\subsection{Implementation}\label{sec:mesa_eos_implementation}

We implemented the calculation of the EoS and the creation of the tables for \mesa as a separate \texttt{Python} module called \texttt{tinyeos}\footnote{\url{https://github.com/tiny-hippo/tinyeos}}. The module uses EoS tables for hydrogen, helium, and a heavy element to model arbitrary mixtures. The currently available tables are \citet{1995ApJS...99..713S}, \citet{2019ApJ...872...51C} and \citet{2021ApJ...917....4C} for hydrogen and helium, and the QEOS \citep{1988PhFl...31.3059M} heavy-element tables for H$_2$O, SiO$_2$, Fe (A. Vazan, priv. comm.) and CO \citep{2023Icar..39415424P}, as well as the SESAME \citep{lyon1978sesame} and AQUA \citep{2020A&A...643A.105H} tables for H$_2$O. The tables are evaluated using the bi-cubic spline \texttt{scipy.interpolate.RectBivariateSpline}, and derivatives are calculated directly from the splines. 

The EoS is available with both $(\log \rho, \log T)$ and $(\log p, \log T)$ as the thermodynamic bases. If a table is only available for one pair of these independent variables, it is inverted by interpolating along isotherms with the piece-wise monotonic cubic spline \texttt{scipy.interpolate.PchipInterpolator} to create tables for the other pair. In addition to the density and temperature or pressure and temperature, the EoS uses the mass fractions of hydrogen and up to three heavy elements as inputs. While the total mixture is calculated during runtime, the three-component heavy-element part of the mixture is pre-calculated and stored as tables.

Using $(\log p, \log T)$ as the thermodynamic basis significantly simplifies the mixture calculation, since each component is assumed to be at the same pressure and temperature. In our module, the default calculation of the $(\log \rho, \log T)$ EoS (the required basis in \mesa) performs a vectorized root-finding procedure with \texttt{scipy.optimize.elementwise} that uses the $(\log p, \log T)$ EoS. This is numerically robust and, due to vectorisation, quite efficient\footnote{While the option exists to calculate the mixture using the density-temperature tables, it is only recommended for specific applications.}.

By default, our EoS builds the tables for \mesa as follows: The EoS is calculated on rectangular grid in $(\log Q, \log T)$, with $\log T_{\rm{(min, max)}} \, [\rm{K}] = 2, 6$ and $\log Q_{\rm{(min, max)}} = -6, 6$ and a resolution of $\Delta \log T \, [\rm{K}] = 0.02$ and $\Delta \log Q = 0.05$. The requirement that the tables be rectangular in $(\log Q, \log T)$ leads to EoS calls with densities that are outside the range of the hydrogen, helium or heavy-element tables. It is therefore left to the responsible user to ensure that the planetary models are within the validity of the respective equations of state (see Figure \ref{fig:eos_logrho_logt}). 

The above procedure was repeated for all possible combinations of the elemental mass fractions with a spacing of $\Delta (X, Y, Z) = 0.1$, and yielded 66 tables. To speed up computation, the building of the tables was parallelized using \textit{joblib.Parallel}. The output of the EoS was checked for bad values, for example, infinities or nans. When that occurred, the bad values were replaced by a distance-weighted nearest-neighbour interpolation. In addition, since the calculation of spline derivatives could  sometimes yield unphysical values, a few crucial quantities, for example, the adiabatic temperature gradient and pressure derivatives, were constrained to remain within certain bounds. For numerical robustness, the resulting tables were smoothed with the distance-weighted nearest-neighbour interpolation on the entire grid. The smoothing could be disabled or repeated; by default, it was done once.

These tables could then be used in \mesa with our custom EoS subroutine. This subroutine contains a few crucial modifications, allowing for the full range of different compositions. Otherwise, it is largely a copy of the original \mesa EoS subroutines and works in the same way. The custom EoS code and tables using the \citet{2021ApJ...917....4C} EoS for hydrogen and helium, and a 50-50 H$_2$O SiO$_2$ mixture from the QEOS tables, is available on GitHub.

\subsection{Example evolution calculations}

In this section, we illustrate the regions of validity of our EoS in the temperature-density space and demonstrate the capabilities of the new EoS. 
We calculated two evolution models of homogeneously mixed $M = 1{\rm{M_J}}$ planets with $T_{\rm{eq}} = 400$ K and bulk metallicities of $Z = 0.012$ and $Z = 0.20$, and proto-solar hydrogen-helium ratios. A 50-50 water-rock mixture represented the heavy elements. We assumed hot-start initial conditions and used the irradiated-grey atmospheric model of \citet{2010A&A...520A..27G} as implemented in \mesa. 

\begin{figure}
    \centering
    \includegraphics[width=\columnwidth]{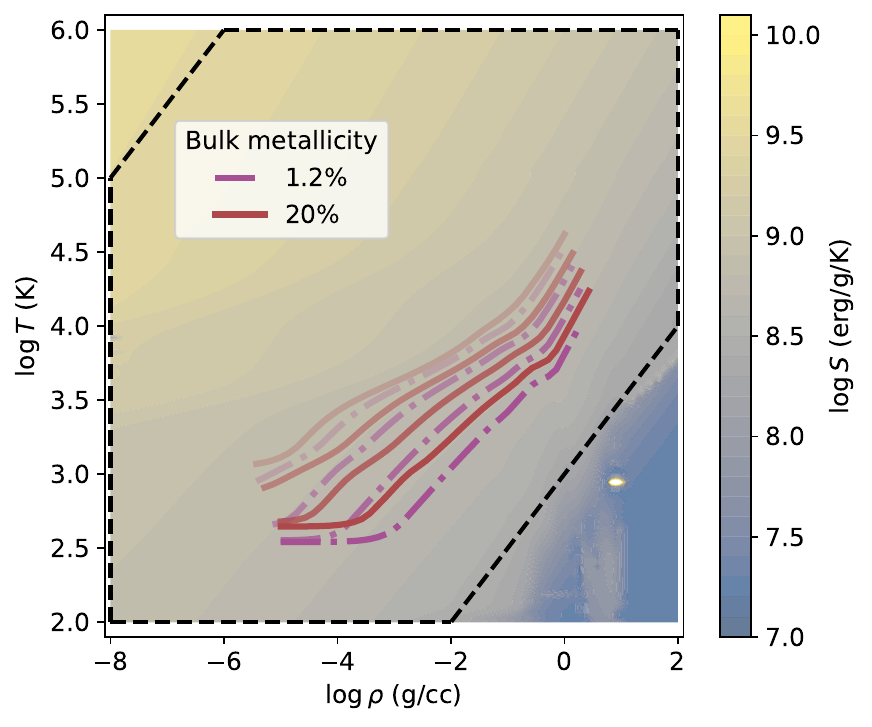}
    \caption{Time evolution of two homogeneous planets in the $(\log \rho, \log T)$ space with $Z = 0.012$ (dash-dotted purple lines) and $Z = 0.20$ (solid red lines) for times between 1 Myr and 4.5 Gyr. More transparent lines correspond to earlier times. The contours show the entropy, and the dashed black lines show the boundaries of the EoS.}
    \label{fig:eos_logrho_logt}
\end{figure}

\begin{figure}
    \centering
    \includegraphics[width=\columnwidth]{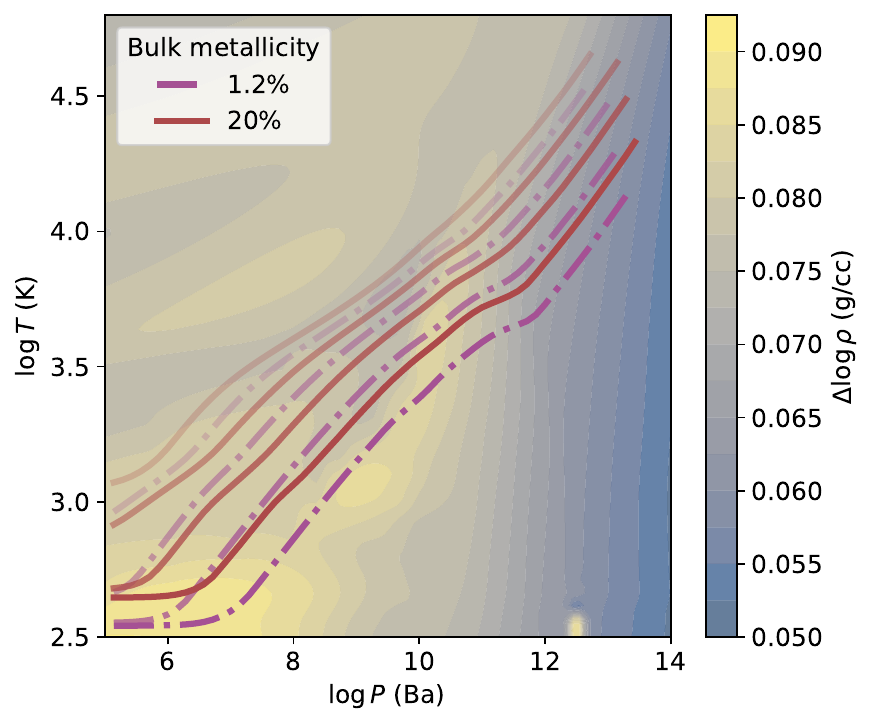}
    \caption{Time evolution of two homogeneous planets in the $(\log p, \log T)$ space with $Z = 0.012$ (dash-dotted purple lines) and $Z = 0.20$ (solid red lines) for times between 1 Myr and 4.5 Gyr. More transparent lines correspond to earlier times. The contours show the difference in density for the two compositions, and the dashed black lines show the boundaries of the EoS.}
    \label{fig:eos_logp_logt}
\end{figure}

The densities and temperatures for times between 1 Myr and 4.5 Gyr are shown in Figure \ref{fig:eos_logrho_logt} for both bulk compositions. The dashed black lines show the boundary of the EoS, and the entropy contours are shown in the background. Figure \ref{fig:eos_logp_logt} shows the same models, but depicts the pressure and temperature profiles instead. Here, the contours show the difference in density $\Delta \log \rho$ due to the different bulk compositions. The higher bulk metallicity leads to a significantly denser and hotter interior compared to a model with solar composition. 

For modeling giant planets, the EoS presented here is a significant improvement over the current options available in \mesa, as it allows for the modeling of planets with arbitrary bulk compositions. This is particularly important since giant planets are often metal-rich and do not share the same composition as their host stars \citep[][]{thorngren_mass-metallicity_2016,teske_metal-rich_2019,muller_towards_2023,howard_giant_2024,2025A&A...693L...4M}.

\section{Opacity}\label{sec:mesa_opacity}

The thermal evolution of giant planets strongly depends on the atmospheric boundary conditions, which regulate the efficiency of energy loss to space. These boundary conditions are governed by the atmospheric temperature-pressure profile and opacities, which control radiative transfer and thus influence the cooling and contraction history of the planet \citep[][]{guillot_cepam_1995, fortney_synthetic_2008}. Accurate treatment of opacities, particularly those of H$_2$, He, and trace species like H$_2$O, CH$_4$, and NH$_3$, is essential for modeling the radiative-convective boundary, which in turn sets the planet's luminosity over time \citep[][]{marley_luminosity_2007, 2008ApJS..174..504F}. Small variations in opacity can lead to significant differences in thermal timescales, highlighting the sensitivity of evolutionary models to atmospheric assumptions.

For giant planets, opacities are also important in the deeper interior. Since giant planets may harbour composition gradients \citep[][]{2024arXiv240705853H}, determining the modes of energy transportation requires knowledge of conductive and radiative opacities. We note that in the deep-interior conditions relevant for giant planets, electron conduction is expected to be significantly more efficient than radiative transport by photons. In \mesa, the conductive opacity is calculated from tabulated thermal conductivities. The tables are an extended version of the calculations from \citet{2007ApJ...661.1094C}, and were privately communicated by A.Y. Potekhin.

For the radiative contribution at high temperatures (beyond a few thousand K), the dominant opacity sources are electron scattering (Thomson scattering) and bound-free and free-free absorption by highly ionized atoms. \mesa primarily uses OPAL opacities \citep[][]{1996ApJ...464..943I} for these regimes. These are extensive tables covering a wide range of temperatures, densities, and chemical compositions, computed using detailed atomic physics calculations.

In low temperatures (mostly relevant for the outer envelope and atmosphere), radiative opacities are hard to determine: At temperatures of $\lesssim 10^4$ K, the opacity becomes dominated by molecules (for example, H$_2$O, CO, TiO), the H$^-$ ion, dust grains, or even clouds. In this regime, \mesa relies on extensive pre-computed tables from specialized codes, such as those by \citet{2005ApJ...623..585F}, \citet{2008ApJS..174..504F,2014ApJS..214...25F}, and AESOPUS \citep{2009A&A...508.1539M,2022ApJ...940..129M,2024ApJ...976...39M}.

While the low-temperature opacities of \citet{2005ApJ...623..585F} include grains, their contribution is absent from the other tables. Additionally, they do not account for potential cloud decks that could locally modify the opacity significantly. To alleviate this, we  also  implemented simple grain and cloud contributions that can be included in the opacity calculation and therefore mimic more complex atmospheric conditions \citep{2019Atmos..10..664P}.

We note that the following methods of modifying the opacities to include additional physics are clearly simplified. However, they provide an accessible way of studying how, and for which parameters, they affect the thermal evolution of giant planets. An alternative and more comprehensive approach would be to use tabulated atmospheric models to determine the boundary conditions for the evolution models. Currently, a few tables exist in \mesa that can be used for stars and brown dwarfs. It would be clearly advantageous to have a variety of tabulated atmospheric models that can be used for irradiated giant planets, and we hope that they can be both constructed and implemented in the future.

\subsection{Grains}

To model refractory grains that stay mixed in the background gas and do not settle into the clouds, we followed the methodology from \citet{2013ApJ...775...10V}. This simple model determines the conditions under which grains are present or absent, and provides a linear fit to the more complex calculations for a grain opacity from \citet{1994ApJ...437..879A}. Namely, for $\log T < \log T_1^* = 0.0245 \log \overline{R} + 3.096$, where $\log \overline{R} = \rho/(10^{-6} \, T)^3$, the grain contribution is supposed to be maximal. This expression deviates from the one in \citet{2013ApJ...775...10V}. There is, in fact, a mistake in the published version, which results in an inconsistent fit to the \citet{1994ApJ...437..879A} grain opacity. Here, we provide the corrected version communicated to us privately by D.~Valencia. For $\log T < \log T_1^*$, the opacity from the grains is 

\begin{equation}
    \log \kappa_{\textrm{grains}} = 0.430 + 1.3143(\log T - 2.85), \,
    \label{eq:grain_opacity}
\end{equation}

\noindent
where $T$ is in Kelvin, and $\kappa_{\textrm{grains}}$ in cm$^2$/g. The total opacity is simply the sum of the gas and grain opacity: $\kappa = \kappa_{\textrm{gas}} + \kappa_{\textrm{grains}}$. For $\log T > \log T_2^* = 0.0245 \log \overline{R} + 3.221$ on the other hand, the grains are thought to have fully evaporated and $\kappa_{\textrm{grains}} = 0$. For  $\log T_1^* < \log T < \log T_2^*$, the grain opacity is linearly interpolated between the two critical values $\kappa_{\textrm{grains}}(\log T_1^*)$ and $\kappa_{\textrm{grains}}(\log T_2^*) = 0$.

The calculations from \citet{1994ApJ...437..879A} assumed that the grain-size distribution is that of the interstellar medium dominated by tiny grains. This results in a very high grain opacity, which is likely not very realistic for evolved giant planets \citep[]{movshovitz_formation_2010}. To account for this, our implementation in \mesa includes a user-defined scaling factor $f_\textrm{grains}$ that is applied to Eq. \ref{eq:grain_opacity}. The \mesa implementation of the grain opacity described here was already used in planet formation calculations \citep{valletta_giant_2020,2024A&A...685A..22M,2025arXiv250701212S} and the characterization of giant exoplanets \citep[][]{delamer_toi-4201_2024}.

\subsection{Clouds}

To account for the effect of clouds on the thermal evolution of giant planets, we implemented a simple model that treats cloud decks as an additional opacity source \citep{2012MNRAS.420...20H,2019Atmos..10..664P,2024MNRAS.529.2242P}. The cloud opacity $\kappa_{\textrm{clouds}}$ is then added to the local gas and grain opacity. The cloud opacity is parametrized as a Gaussian function:

\begin{equation}
    \kappa_{\textrm{clouds}} = \kappa_{\rm{c}} \cdot \exp\left[-\delta_c \left(1 - p / p_c\right)^2\right], 
    \label{eq:cloud_opacity}
\end{equation}

\noindent
where $\delta_c$ and $p_c$ are the cloud-deck thickness and location, and $\kappa_{\rm{clouds}}$ is the cloud-opacity normalisation. This yields a cloud opacity that is maximal at $p_c$ and symmetrically spread around it with some width defined by $\delta_c$. This formalism can easily be generalized to account for several cloud decks, in which case the total opacity is $\kappa_{\textrm{clouds}} = \sum_i\kappa_{\textrm{clouds, i}}$.

The three free parameters determining the cloud opacity are poorly constrained and depend on the molecules making up the cloud deck. Reasonable estimates for the three parameters are $\kappa_{\rm{clouds}, 0} \sim 0.05 - 1.0$ cm$^2$/g, $\delta_c \sim 10 - 100$, and $p_c \sim 10^5 - 10^7$ Ba{\footnote{We remind the reader that 1 Barye equals 0.1 Pascal.} \citep{2024MNRAS.529.2242P}. For generality and simplicity, in our current implementation, the cloud-deck location is constant in time. However, we note that \citet{2024MNRAS.529.2242P} suggested an evolving cloud-deck location that matches pre-calculated condensation curves of the cloud species of interest \citep{2006ApJ...648.1181V,2010ApJ...716.1060V}. As the planet cools, the condensation curves intersect the atmospheric pressure-temperature profile at different pressures, adjusting the position of the cloud deck.

\subsection{Opacity windows}
Recently, \citet{2024ApJ...967....7M} investigated models of Jupiter in which the radiative opacity decreases significantly around $T \sim 2000\,\mathrm{K}$, caused by increased hydrogen transparency and potential depletion of alkali metals. This opacity window can lead to the formation of a deep radiative zone in the planetary envelope, typically between $1$ and $10\,\mathrm{kbar}$. The dip in the radiative opacity $\kappa$ was modeled as a simple Gaussian reduction:

\begin{equation}
    \kappa = \kappa_0 \left(1 - \alpha \, \textrm{e}{^{(-0.5 (\log T - \mu) / \sigma)^2}}\right) \, ,
    \label{eq:opacity_scaling}
\end{equation}

\noindent
where $\kappa_0$ is the unmodified opacity from \citet{2014ApJS..214...25F}, and $\log T$ is the logarithm of the local temperature in K. The parameters $\mu$, $\sigma$ and $\alpha$ determine the location, width and depth of the opacity reduction. The nominal values of $\mu = 3.3$ and $\sigma = 0.15$, and $\alpha = 0.9$ yield a qualitatively good match to the location and width of the opacity reduction from \citet{1994Icar..112..354G}.

While detailed opacity calculations exist that can account for a depletion in alkali metals \citep{2024ApJ...976...39M,2025A&A...693A.308S}, there are large uncertainties related to the depletion factor and individual opacities. The simple approach from \citet{2024ApJ...967....7M} yields similar results, allowing for easy investigation of how different parameter values influence the evolution.

The radiative region suppresses convective mixing and thus decouples the outer atmospheric layers from the planet’s deep interior. They therefore argued that late-stage accretion of heavy elements can enrich the planetary atmosphere without significantly altering the deeper composition. It was shown that such a configuration is stable over gigayear timescales. This mechanism resolves the apparent inconsistency between observed atmospheric enrichment of Jupiter and predictions from homogeneous interior models, and has important implications for interpreting atmospheric compositions in both solar-system and extrasolar gas giants. \newline

To demonstrate the modifications of the opacity, we calculated an evolution model of a $1{\rm{M_J}}$ planet with $T_\textrm{eq} = 200$ K using the \citet{2014ApJS..214...25F} opacity as implemented in \mesa. Then, we used the final model at about 4.5 Gyr to calculate the opacity profiles, including grains, clouds, or an opacity window.

\begin{figure}
    \centering
    \includegraphics[width=\columnwidth]{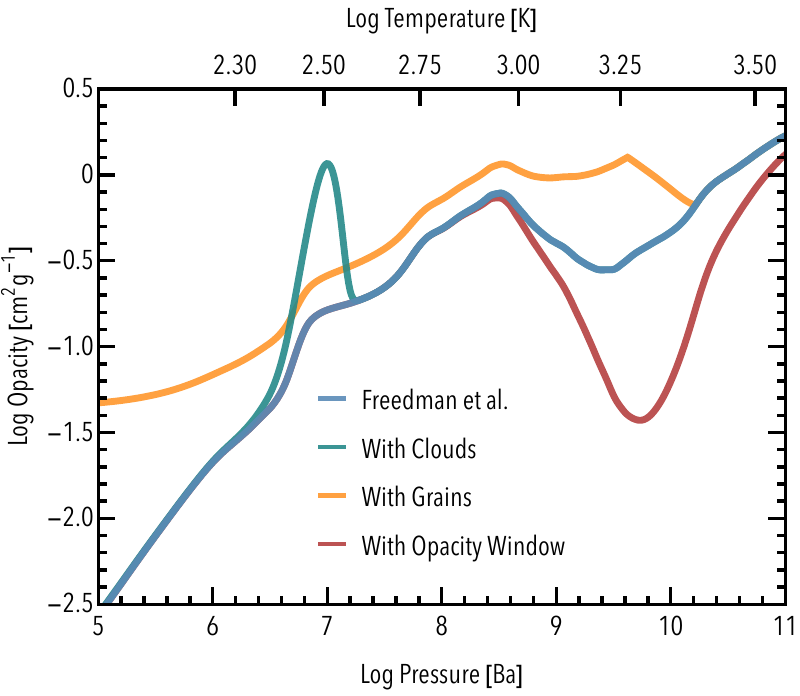}
    \caption{Opacity profile of a $M = 1{\rm{M_J}}$ planet with $T_\textrm{eq} = 200$ K at about 4.5 Gyr. The pressure-temperature conditions were from the baseline model using the Freedman et al. opacity (blue line). The orange line includes grains with a scaling factor $f_\textrm{grains} = 0.1$. The effect of a single optically thick cloud deck at 10 bar with $\delta_c = 10$ and $\kappa_{\rm{clouds}, 0} =  1$ cm$^2$/g is shown in the green line. The dashed purple line illustrates how an opacity window could be created due to a lack of alkali metals.}
    \label{fig:kap_logp_profile}
\end{figure}

The results are shown in Figure \ref{fig:kap_logp_profile}. The opacity changes significantly by these modifications. The increased opacity due to grains or clouds would slow the cooling, yielding a different planetary radius and luminosity at a given time \citep[][]{vazan_effect_2013,2024MNRAS.529.2242P}. The opacity window would not only modify the cooling rate but also add a deep radiative zone that could disconnect the deeper envelope from the atmosphere \citep{1994Icar..112..354G,howard_exploring_2023,2024ApJ...967....7M}. All of these have implications for the characterization of planetary interiors and atmospheres, and therefore should be considered when modeling giant planets.

%%%%%%%%%%%%%%%%%%%%%%%%%%%%%%%%%% Convective Mixing %%%%%%%%%%%%%%%%%%%%%%%%%%%%%%%
\section{Convective mixing}\label{sec:mixing}
Convective mixing also plays a vital role in giant planet evolution as it redistributes material throughout a planet, thereby directly altering its internal structure \citep[][]{vazan_effect_2013, 2020A&A...638A.121M, 2024ApJ...971..104S, Knierim2025}. This process requires energy (or releases it in the case of sedimentation) and modifies material properties (for example, the EoS or opacity), substantially changing the evolutionary trajectory. Accurately modeling giant planets therefore requires us to account for this fundamental process. 
\par
Since mixing plays an equally important role in stellar evolution, \mesa already incorporates various convective mixing algorithms. In general, convective mixing is modeled as a diffusive process, with the diffusion coefficient determined by mixing length theory. Convective stability is evaluated using the Schwarzschild (or Ledoux) criterion. In its most basic form, this evaluation is done by computing the difference $y = \nabla_\mathrm{rad}-\nabla_\mathrm{ad}$ (or $\nabla_\mathrm{L}$ instead of $\nabla_\mathrm{ad}$ in the case of the Ledoux criterion), and then locating the cell where $y$ changes its sign. However, in the presence of discontinuous composition gradients, this ``sign-change'' approach produces convective boundaries where $\nabla_\mathrm{rad} \neq \nabla_\mathrm{ad}$ on the convective side--failing to satisfy the Schwarzschild criterion \citep{Gabriel2014}. Hence, this approach can underestimate the extent of convective regions, yielding unreliable models. To address this issue, \mesa first introduced the \texttt{predictive\_mixing} algorithm, and later improved upon it with the \cpm algorithm.
The general idea of the \cpm algorithm is to expand convective regions until $\nabla_\mathrm{rad} = \nabla_\mathrm{ad}$ on the convective side of the boundary.
It does this by tentatively treating adjacent radiative cells as convective, fully mixing this expanded region, and re-evaluating the Schwarzschild criterion. If the newly added cell remains radiative, the algorithm reverts to the original convective region; if it becomes convective, the expansion continues \citep[for details, see][]{2023ApJS..265...15J}.
\par
Although these algorithms perform well for stars, they tend to become numerically unstable when applied to giant planets.
The primary reason is the more complex EoS. While most stars can be reasonably described by a fully ionized ideal gas, the planetary EoS is significantly more intricate (see Sect.~\ref{sec:mesa_eos}).
Under planetary conditions, small changes in composition can induce substantial variations in the overall structure. Moreover, the derivatives of the EoS often behave poorly, particularly near phase transitions.
These issues result in ill-conditioned Jacobians, which can cause the Henyey solver to fail.
\par
To address these instabilities, \citet{Knierim2024} (hereafter KH24) introduced the \gm algorithm. This method monitors changes in the composition profile between steps. If the change exceeds a certain threshold, the algorithm reduces the mixing efficiency and shortens the timestep.

At its core, the \gm algorithm monitors changes in a cell's composition and the model's overall heterogeneity:
\begin{align}
    h_i^2 = \int_{0}^{M} (X_i' - X_i)^2 \frac{\mathrm{d}m}{M},
\end{align}
where $X_i$ and $X_i'$ denote the mass fractions of species $i$ before and after a step, respectively. If either the abundance change or the heterogeneity $h_i$ exceeds a user-defined threshold, \mesa's \texttt{mix\_factor} and the time step are reduced, and a retry is initiated.
The algorithm provides several options to customise this behavior, all of which are documented in the \texttt{controls.defaults} file in the GitHub repository.
\par
Most notably, it includes an adaptive damping mechanism that tracks the number of retries and re-dos performed by the solver. If a retry was required in the previous step, the algorithm increases \texttt{mix\_factor} gradually from its last (damped) value rather than resetting to the original value. If subsequent steps succeed, \texttt{mix\_factor} is incrementally restored to its reference value. 
Conversely, if retries persist and \texttt{mix\_factor} drops below a user-defined minimum, mixing is temporarily disabled to facilitate solver convergence. Similarly, the time step is not reduced indefinitely but only down to a specified lower limit.
The \gm algorithm also allows for asymmetric $h_i$ thresholds, distinguishing between high- and low-abundance regions. For example, high-$Z$ cells, typically less stable under planetary conditions, can be treated with more stringent thresholds.
\par
Many of these modifications were also applied to the \cpm algorithm. A key difference is that $h_i$ can be evaluated during \cpm's tentative region expansion, allowing the process to be aborted early if the heterogeneity becomes too large.
To support these new capabilities, the \texttt{do\_eos\_for\_cell} routine in \texttt{star/private/micro.f90} was modified, and a new routine, \texttt{solve\_eos\_given\_PgasS}, was added to \texttt{eos/eos\_support.f90}.
We note that the damping of the mixing is acceptable as long as the mixing timescale remains much shorter than the Kelvin–Helmholtz timescale—that is, as long as mixing continues to occur significantly faster than the planet’s thermal evolution. Since the convective mixing timescale in gas giants is typically six to seven orders of magnitude shorter than the Kelvin–Helmholtz timescale, we can afford to damp the mixing somewhat without affecting the overall evolution.
Figure \ref{fig:gentle_mixing_visualization} illustrates this process.
\begin{figure}
\centering
  \includegraphics[width=0.9\columnwidth]{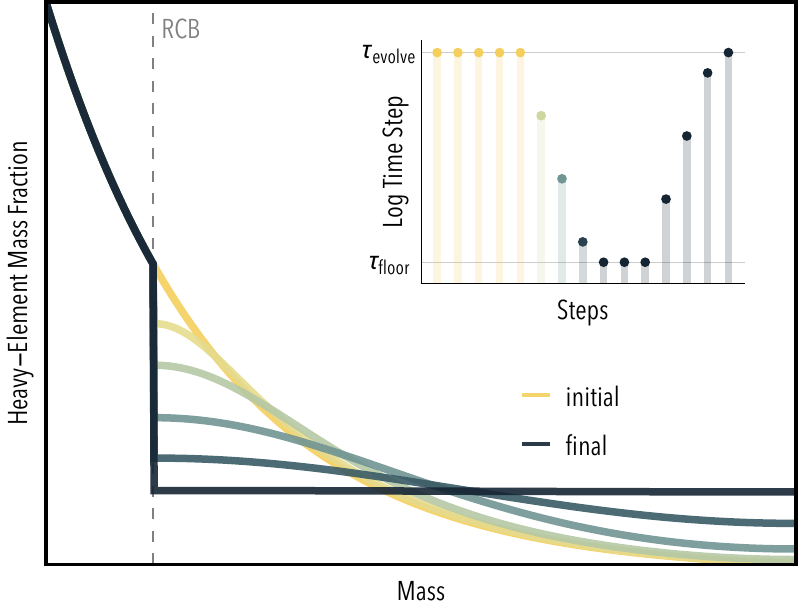}
  \caption{Sketch of the \gm algorithm. Rather than transitioning directly from the initial to the final heavy-element profile, the algorithm inserts several intermediate steps. The inset in the top right corner shows how the damping of mixing efficiency is accompanied by a reduction in time step.}
  \label{fig:gentle_mixing_visualization}
\end{figure}
As a result, \mesa mixes unstable regions over several smaller time steps instead of a single, large one.
In addition, \gm also monitors changes during \cpm: If the newly expanded region alters the composition profile beyond the allowed threshold, \gm reverts the expansion and exits \cpm—again accompanied by a reduction in the time step.
\begin{figure}
\centering
   \includegraphics[width=\columnwidth]{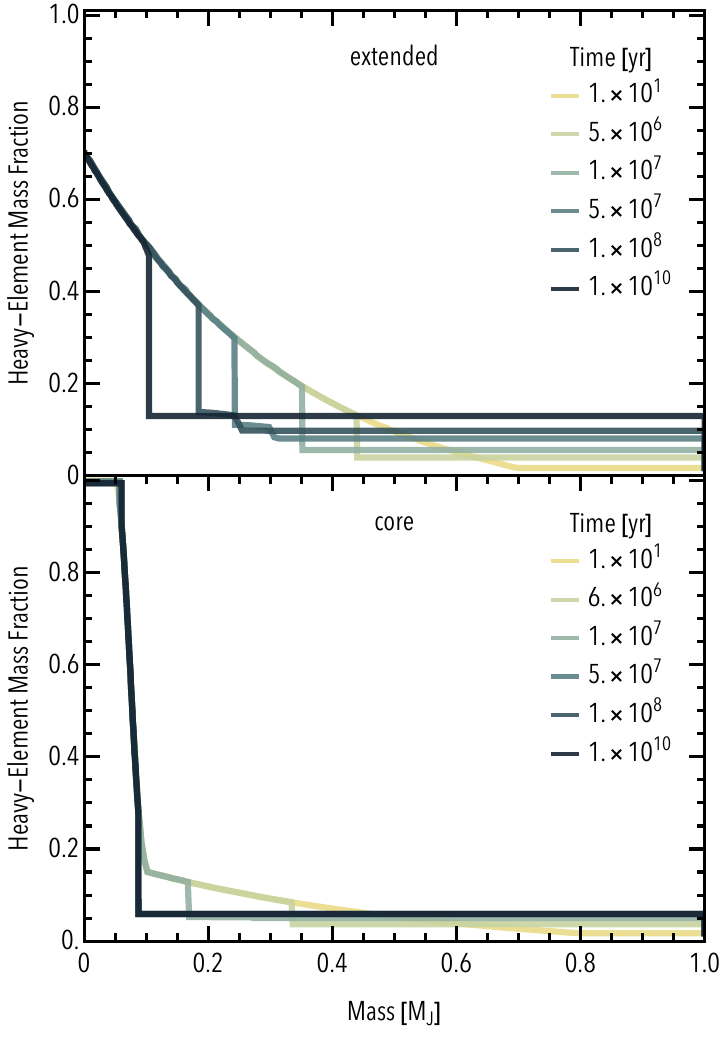}
     \caption{Evolution of the composition profile for a \SI{1}{\MJ} planet and an entropy of \SI{8}{\kbperbary}, starting from an extended composition profile (top) and a core-like structure (bottom).}
     \label{fig:Z_evolution}
\end{figure}

Furthermore, we extended \cpm to better resolve convective boundaries under planetary conditions. In its default implementation, \cpm holds either pressure and temperature or density and temperature constant during the tentative expansion of a convective region. However, this can introduce significant errors, as temperature may vary strongly when mixing material in planetary interiors. To address this, we introduced two new modes that instead hold either pressure and density or pressure and total entropy in the convective region constant. While fixing any such pair of variables is, strictly speaking, unphysical, the pressure–entropy mode in particular provides substantially greater numerical stability than the default approach.
Since its original release in KH24, \gm has undergone several improvements. Most notably, it now includes an adaptive scheme that attempts to mix as much as possible without failures.
%\par
Figure \ref{fig:Z_evolution} shows the evolution of two different initial composition profiles utilizing \gm. 
The extended profile (left panel) erodes significantly over \SI{10}{\Gyr}, leading to substantial enrichment of the envelope. On the other hand, the core-like profile (right panel) is more resilient, losing only its outer gradient during the same period.
Despite the differing degrees of mixing, both models exhibit significant changes to their internal structures.
Figure \ref{fig:mixing_temperature_comparison} in Appendix \ref{sec:appendix_gentle_mixing} compares the resulting temperature profiles for simulations with and without mixing after \SI{10}{\Gyr}. If mixing is ignored, composition gradients artificially suppress convection, resulting in interiors that are considerably hotter than they should be. When mixing is included, only the innermost physically stable layers retain elevated temperatures; the outer interior approximately recovers the temperature structure of a homogeneous model with the same bulk composition.
Thus, although the extent of convective mixing depends on the details of the entropy and composition profiles (see KH24 for a full theoretical treatment), accurately modeling the mixing process is crucial to obtain realistic internal structures.

\section{Helium rain}\label{sec:helium}
While hydrogen and helium can be mixed, under some pressures and temperatures, helium becomes immiscible with hydrogen, leading to phase separation. This process is known as ``helium rain" \citep{Stevenson_1977b}. The settling of helium leads to energy release and can affect planetary evolution.  
Although the H-He phase diagram is still being investigated and the exact conditions for H-He demixing remain uncertain \citep[][]{Lorenzen2011,Schoettler2018,Brygoo2021}, it is clear that helium rain has occurred in both Jupiter and Saturn \citep[for example,][and references therein]{Fortney2004,Howard2024}.
Therefore, we also incorporated helium rain and element sedimentation more generally in \mesa \citep[see][for details]{Knierim2025}. The implementation proceeds in three steps: (1) determine the immiscibility region using a user-supplied phase diagram, (2) remove the sedimenting species, and (3) repeat steps (1) and (2) until convergence is reached.
\par

In step (1), the user-supplied phase diagram is read into a custom \texttt{phase\_diagram} type. This type stores the lower and upper miscible number fractions, together with the corresponding temperature and pressure values. The user can specify the elements to be included and to apply a temperature shift.
To determine whether a cell undergoes precipitation, we first linearly interpolate the lower and upper miscible fractions in pressure–temperature space. We then evaluate whether the local abundance ratio lies above, within, or below the miscibility gap. For ratios inside the gap, we calculate the depletion required for the composition to reach the lower miscibility boundary.
The resulting abundance change is converted into a mass abundance difference and stored in the array \texttt{dXi}. We compute the changes in the other mass fractions assuming that their ratios remain constant throughout the sedimentation process. We note that in fact it is not trivial to determine the other mass fractions throughout the evolution (see Appendix \ref{sec:helium_fraction}).
\par

At this stage, \texttt{dXi} can be further modified by user-defined constraints. These include limiting the maximum abundance change per step, enabling many small adjustments rather than a single large rain-out event. Additional options allow one to close gaps within a precipitating region and enforce monotonicity of the composition gradient.
The algorithm also accounts for core and ocean formation when the innermost cell undergoes precipitation. In this case, since there is no lower cell to transfer material to, \texttt{dXi} is adjusted to enrich the bottom cell up to the upper miscibility boundary. The process continues outward until all immiscible cells are resolved.
\par
In step (2), the specified \texttt{dXi} values are subtracted from the sedimenting species, 
using one of three modes, listed here in order of increasing complexity.
In \texttt{instant\_rain}, the code evaluates the stability of the entire model and transfers the sedimenting element in a single step to the deepest miscible layer. The thermal structure is then updated globally. 
The \texttt{bottom\_up\_rain} mode proceeds cell by cell, starting from the deepest layer and moving upward, transferring material and updating the thermal structure incrementally.
Finally, the \texttt{advection\_diffusion\_rain} mode solves the advection-diffusion equation, modeling element sedimentation as an advective process and convection as diffusion, with \mesa's \texttt{D\_\rm{mix}} used as the diffusion coefficient. Between \mesa time steps, the solver iterates until a convergent solution is reached. This mode introduces two additional free parameters: (1) the sedimentation velocity of the element and (2) the time step used by the advection-diffusion solver.
\par

In \texttt{advection\_diffusion\_rain}, the algorithm solves the standard advection-diffusion equation for an isotropic sphere:

\begin{align}
    \diffp{Y}{t} + \diffp{}{m} \left[4 \pi r^2 \rho \left(-D_{\rm{mix}} \diffp{Y}{r} + v_{\rm{sed}} Y\right)\right] = 0 \, ,
\end{align}

\noindent
where $Y$ is the helium mass fraction, $t$, $m$ and $r$ the time, mass, and radial coordinates, $D_{\rm{mix}}$ is the convection diffusion coefficient from mixing-length theory, and $v_{\rm{sed}}$ is the sedimentation velocity. In the case of helium rain, the sedimentation velocity is only nonzero if helium is actively raining out according to the phase diagram. The equation is solved with the implicit backward-Euler scheme on a non-uniform grid. The method contains two additional free parameters: (1) the time step used by the solver and (2) the velocity of the sediment species.
For all methods, the energy of the sedimentation is accounted for in the energy equation and \mesa's output quantities (for example, the total energy). 
Figure~\ref{fig:helium_rain} compares the evolution of a homogeneous Saturn-mass planet across the different helium rain modes.
\begin{figure}
\centering
\includegraphics[width=\columnwidth]{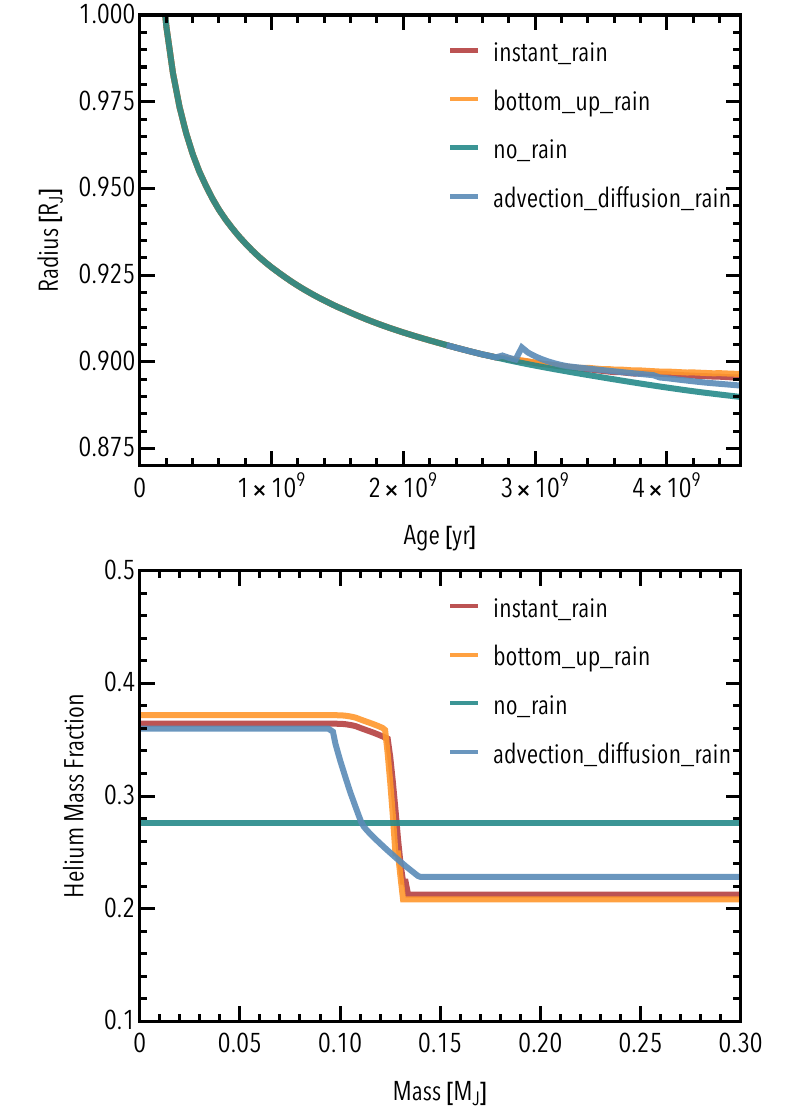}
   \caption{Comparison of different helium rain implementations, showing the radius evolution (top) and the final helium mass fraction profile (bottom) of a homogeneous Saturn-mass model.}
   \label{fig:helium_rain}
\end{figure}
The \texttt{instant\_rain} and \texttt{bottom\_up\_rain} methods yield very similar radius evolutions and helium distributions. By contrast, the advection-diffusion solver produces a smoother core-envelope transition, shaped by the ratio of advection to diffusion. A higher sedimentation velocity results in a steeper gradient and transports more helium into the deep interior. In the case shown, we adopt a sedimentation velocity $v_{\rm{sed}} =$ \SI{0.825}{\cm \per \s}, though both higher and lower values are plausible given the current uncertainty in the microphysics of helium rain.
\par
These modes provide users with significant flexibility. In light of the uncertainties surrounding element sedimentation, the implementation also supports options such as applying a temperature shift to the phase diagram to better match specific observational constraints.

The process of Helium rain is largely implemented in a new file, \texttt{element\_sedimentation.f90}, located in the \texttt{star/private} directory. The core routine, \texttt{do\_element\_sedimentation}, is called during the evolve loop, immediately before \mesa's diffusion solver.
%The code is available at \textbf{link-to-github-repository}.

\section{Comparison to other evolution models}
Currently, \mesa is the only open source Henyey code that is capable of simulating the full thermal and compositional evolution of giant planets. This makes direct comparisons with other studies rather challenging. Not only are alternative codes inaccessible to us, but they also often use different EoSs (and the numerical details of their implementation) and atmospheric models.

Under these constraints, we first tackled the seemingly simple case of a homogeneously mixed typical warm Jupiter with $M = 1 M_{\rm{Jup}}$ at $T_{\rm{eq}} = 500$ K, and with zero or 5\% bulk metallicity. However, as discussed below, even this comparison is not trivial.

We compared the following models:

\begin{itemize}
    \item Two models calculated with \mesa using the $T-\tau$ and irradiated grey \citep{2010A&A...520A..27G} atmospheric models with the \citet{2014ApJS..214...25F} gas opacities, the \citet{2021ApJ...917....4C} hydrogen-helium EoS, and the QEOS $H_2O$ for the heavy elements.
    \item One model calculated with the open-source code \texttt{GASTLI} \citep{2024A&A...688A..60A}, which uses its own grid of atmospheric models, the hydrogen-helium equation of state from \citet{2019ApJ...872...51C} combined with the one from \citet{2023A&A...672L...1H}, and the AQUA equation of state \citep{2020A&A...643A.105H} for the heavy elements. We note that \texttt{GASTLI} is not a full Henyey code and is incapable of treating transport of energy or chemical elements. It calculates static models for different internal temperatures, and then derives the evolution from that.
    \item One model calculated with the closed-source code \texttt{CEPAM} \citep{guillot_cepam_1995} shared privately with us by S.~Howard. The models used the non-grey atmospheric model from \citet{2015A&A...574A..35P}, the hydrogen-helium equation of state from \citet{2019ApJ...872...51C} combined with the one from \citet{2023A&A...672L...1H}, and the SESAME water equation of state \citep{lyon1978sesame} for the heavy elements. While \texttt{CEPAM} is a Henyey code, it currently lacks the ability to model the transport of chemical elements.
\end{itemize}

Figure  \ref{fig:radius_comparison} shows the radius evolution for both zero and 5\% metallicity. We find that after a few Gyr, for $Z = 0$ all models agree on the planetary radius within about 2\%, while for the $Z = 0.05$ case the results diverge more significantly, with the smallest and largest radii differing by about 5\%. For the pure hydrogen-helium case, the EoSs used in the models are similar. Therefore, the observed variation is probably due to the different atmospheric models. For the $Z = 0.05$ case, there is an additional complication that all models use a different EoS for the heavy elements. Compared to many recently reported observational errors on the radii of giant planets, these differences between the models are rather large and would also translate into differences in the inferred bulk compositions. Given the expected accuracy of radius measurements (on the order of a few percent), theoretical uncertainties are important and should be considered \citep{2020ApJ...903..147M,howard_giant_2024}. 

We also note that the evolution of the planetary radius is typically shown in units of Jupiter's radius. However, not all models use the same radius for Jupiter: While planetary scientists commonly use the volumetric mean radius of Jupiter ($6.9911 \times 10^9$ cm), in exoplanetary sciences Jupiter's equatorial radius ($7.1492 \times 10^9$ cm) is often used. This leads to a non-negligible difference that can affect results and conclusions drawn from models. For this comparison, for example, we had to renormalise the results from \texttt{GASTLI}, since it uses Jupiter's equatorial radius.

We next present a rough comparison of the mixing and helium rain modules. For that, we reconstructed the Jupiter model presented by  \citet{Sur2025} in \mesa. Specifically, we adopted a pure-water EoS for the heavy elements and matched the initial temperature and composition profile. For helium rain, we used the \texttt{instant\_rain} algorithm together with the phase diagram of \citet{Schoettler2018}, shifted by \SI{370}{\K}. In addition, we compare our model to that presented in \citet{Howard2024}, where Jupiter's interior was modeled with a heavy-element core surrounded by a H–He envelope. The results are presented in Figure~\ref{fig:helium_rain_comparison}.

\begin{figure}
    \centering
    \includegraphics[width=0.9\columnwidth]{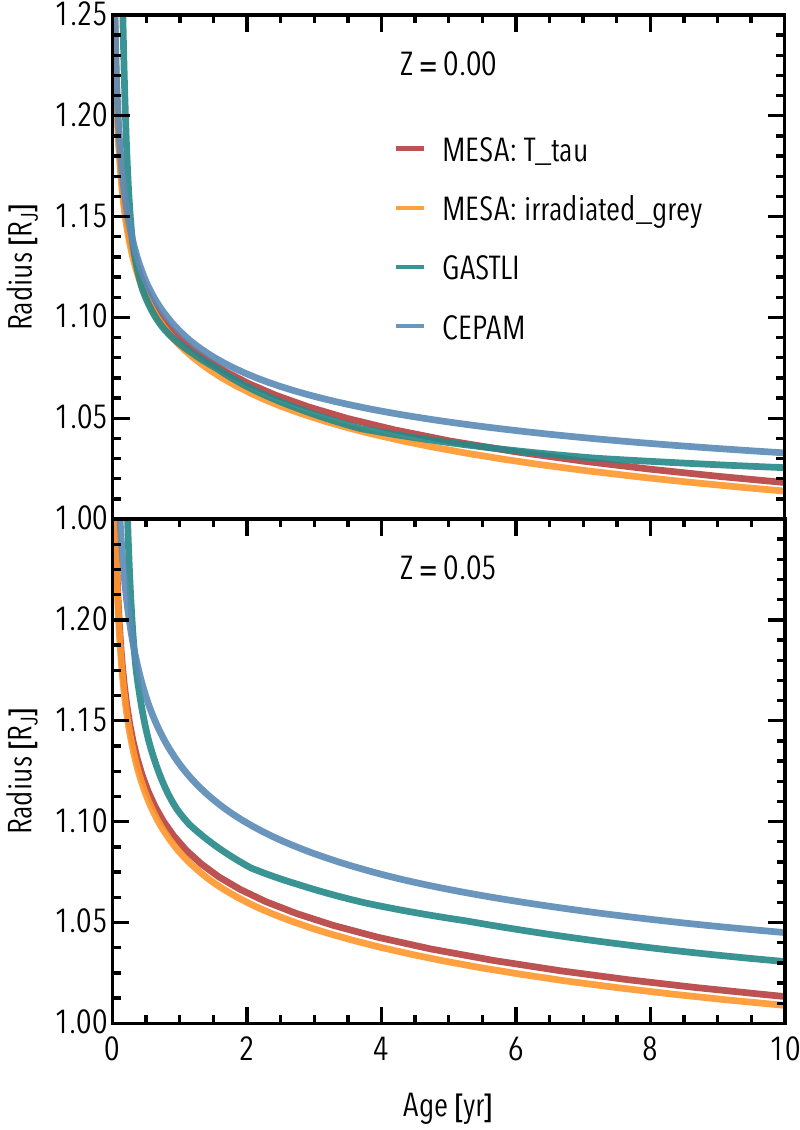}
    \caption{Radius evolution for a $1{\rm{M_J}}$ planet with $T_{\rm{eq}} = 500$ K for zero (top) and 5\% bulk metallicity (bottom). The lines correspond to different models: Orange and red for \mesa with the $T-\tau$ and irradiated grey atmospheric models, green for \texttt{GASTLI}, and blue for \texttt{CEPAM}.}
    \label{fig:radius_comparison}
\end{figure}

%%%%%%%%
\begin{figure}
\centering
\includegraphics[width=0.9\columnwidth]{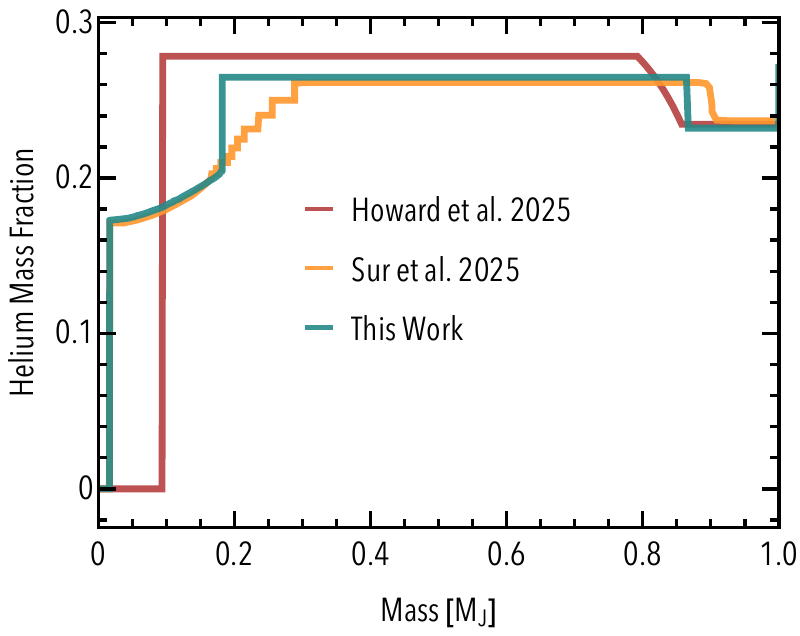}
\caption{Distribution of helium (mass fraction) in present-day Jupiter. Shown are the results inferred by this work,  \citet{Sur2025}, and \citet{Howard2024}.}
\label{fig:helium_rain_comparison}
\end{figure}
%%%%%%%%

We find that all three models reproduce Jupiter’s present-day helium abundance. This agreement is expected, since each study applied a temperature shift to its phase diagram to achieve consistency with the measured helium abundance in Jupiter's atmosphere. Although the detailed numerical values differ slightly, as expected given the varying assumptions, the qualitative trends are similar. 

In addition, the degree of mixing in the dilute core region is similar in our model and the one of \citet{Sur2025}. However, our simulation produces a smooth profile of the helium fraction in the deep interior while simultaneously maintaining a sharp boundary with the envelope. Also, our model does not include convective ``staircases", namely, convective regions of uniform composition separated by sharp composition gradients. In the absence of semi-convection, such staircases are not expected to form and are a numerical artifact \cite[see][for discussion]{Vazan2018}. Crucially, our model using \gm erodes much more of the outer dilute core than the model by \citet{Sur2025}, which employs the sign-change approach to determine convective boundaries (see discussion in Sec. \ref{sec:mixing}).  In general, we find that the different codes produce similar interior profiles, but the exact shapes of the curves depend on the numerical details.

Overall, our results demonstrate that the different evolution models yield similar results, but also that a direct comparison is extremely difficult given all the details and subtleties that go into these models. It is clearly desirable to have detailed model comparisons and benchmarks, but this is beyond the scope of this paper. We note, however, that there is currently an effort underway within the  consortium of the PLATO mission \citep{2025ExA....59...26R} under work-package  \texttt{WP116100} to compare (and in the longer future, benchmark) planetary evolution models. 

\section{Summary}\label{sec:conclusions}
We have introduced several modifications to the \mesa code relevant for giant planets. These include a new equation of state, various treatments of the radiative opacity, and the processes of convective mixing and settling. As the \mesa code is public, our modifications  can be used by the community, paving the way for deeper insights into the formation and characteristics of giant planets in the Solar System and those orbiting around other stars.  

Clearly, planetary evolution models can be further improved. Such improvements include more realistic atmospheric models, conductivities, and other thermodynamic properties. In addition, it would be valuable to benchmark the different evolution models in order to validate them and investigate how the simulation approach and numerical details affect the results.

We hope to address these topics in future research. Indeed, to take full advantage of the current and upcoming observational data, there is a need to enhance the capabilities for planetary evolution studies and push our understanding of these planetary objects to the next level. 

\section*{Data availability}
Our planetary \mesa version can be found in the GitHub repository  \url{https://github.com/uzhplanets/mespa}. 

\begin{acknowledgements}
    We thank the anonymous referee and the editor for valuable comments. 
    We thank Andrew Cumming for guidance and valuable discussions, Sho Shibata and Mark Eberlein for contributions to the custom equation of state subroutines, and Allona Vazan for sharing the QEOS tables of H$_2$O, SiO$_2$ and Fe. 
    We acknowledge support from the Swiss National Science Foundation (SNSF) grant \texttt{\detokenize{200020_215634}} and the National Centre for Competence in Research ‘PlanetS’ supported by SNSF.
    Extensive use was also made of the Python packages \texttt{NumPy} \citep{harris2020array}, \texttt{SciPy} \citep{mckinney-proc-scipy-2010,2020SciPy-NMeth}, \texttt{Matplotlib} \citep{Hunter2007}, \texttt{Jupyter} \citep{jupyter}, \texttt{fortranformat} \citep{py_fortranformat}, \texttt{joblib} \citep{joblib}, \texttt{mesa\_reader} \citep{py_mesa_reader}, \texttt{tinyeos} \citep{tinyeos}, \texttt{mesatools} \citep{pymesatools}, \texttt{py\_mesa\_helper} \citep{pyMesaHelper}, and the Fortran code \texttt{mesa\_custom\_eos}
    \citep{mesaCustomEoS} based on the \mesa EoS subroutines.  
\end{acknowledgements}

\bibliographystyle{aa}
\bibliography{library}

\begin{thebibliography}{89}
\expandafter\ifx\csname natexlab\endcsname\relax\def\natexlab#1{#1}\fi

\bibitem[{{Acu{\~n}a} {et~al.}(2024){Acu{\~n}a}, {Kreidberg}, {Zhai}, \& {Molli{\`e}re}}]{2024A&A...688A..60A}
{Acu{\~n}a}, L., {Kreidberg}, L., {Zhai}, M., \& {Molli{\`e}re}, P. 2024, \aap, 688, A60

\bibitem[{{Alexander} \& {Ferguson}(1994)}]{1994ApJ...437..879A}
{Alexander}, D.~R. \& {Ferguson}, J.~W. 1994, \apj, 437, 879

\bibitem[{{Antia}(1993)}]{1993ApJS...84..101A}
{Antia}, H.~M. 1993, \apjs, 84, 101

\bibitem[{Arnold(2025)}]{py_fortranformat}
Arnold, B. 2025, fortranformat, \url{https://github.com/brendanarnold/py-fortranformat}

\bibitem[{{Baraffe} {et~al.}(2008){Baraffe}, {Chabrier}, \& {Barman}}]{2008A&A...482..315B}
{Baraffe}, I., {Chabrier}, G., \& {Barman}, T. 2008, \aap, 482, 315

\bibitem[{{Brygoo} {et~al.}(2021){Brygoo}, {Loubeyre}, {Millot}, {Rygg}, {Celliers}, {Eggert}, {Jeanloz}, \& {Collins}}]{Brygoo2021}
{Brygoo}, S., {Loubeyre}, P., {Millot}, M., {et~al.} 2021, \nat, 593, 517

\bibitem[{{Burrows} {et~al.}(2001){Burrows}, {Hubbard}, {Lunine}, \& {Liebert}}]{2001RvMP...73..719B}
{Burrows}, A., {Hubbard}, W.~B., {Lunine}, J.~I., \& {Liebert}, J. 2001, Reviews of Modern Physics, 73, 719

\bibitem[{{Cassisi} {et~al.}(2007){Cassisi}, {Potekhin}, {Pietrinferni}, {Catelan}, \& {Salaris}}]{2007ApJ...661.1094C}
{Cassisi}, S., {Potekhin}, A.~Y., {Pietrinferni}, A., {Catelan}, M., \& {Salaris}, M. 2007, \apj, 661, 1094

\bibitem[{{Chabrier} \& {Debras}(2021)}]{2021ApJ...917....4C}
{Chabrier}, G. \& {Debras}, F. 2021, \apj, 917, 4

\bibitem[{{Chabrier} {et~al.}(2019){Chabrier}, {Mazevet}, \& {Soubiran}}]{2019ApJ...872...51C}
{Chabrier}, G., {Mazevet}, S., \& {Soubiran}, F. 2019, \apj, 872, 51

\bibitem[{Delamer {et~al.}(2024)Delamer, Kanodia, Cañas, Müller, Helled, Lin, Libby-Roberts, Gupta, Mahadevan, Teske, Butler, Yee, Crane, Shectman, Osip, Beletsky, Monson, Hebb, Powers, Wisniewski, Alvarado-Montes, Bender, Dong, Han, Ninan, Robertson, Roy, Schwab, Stefánsson, \& Wright}]{delamer_toi-4201_2024}
Delamer, M., Kanodia, S., Cañas, C.~I., {et~al.} 2024, {\textbackslash}apjl, 962, L22, \_eprint: 2307.06880

\bibitem[{{Ferguson} {et~al.}(2005){Ferguson}, {Alexander}, {Allard}, {Barman}, {Bodnarik}, {Hauschildt}, {Heffner-Wong}, \& {Tamanai}}]{2005ApJ...623..585F}
{Ferguson}, J.~W., {Alexander}, D.~R., {Allard}, F., {et~al.} 2005, \apj, 623, 585

\bibitem[{{Fortney} \& {Hubbard}(2004)}]{Fortney2004}
{Fortney}, J.~J. \& {Hubbard}, W.~B. 2004, \apj, 608, 1039

\bibitem[{Fortney {et~al.}(2008)Fortney, Marley, Saumon, \& Lodders}]{fortney_synthetic_2008}
Fortney, J.~J., Marley, M.~S., Saumon, D., \& Lodders, K. 2008, {\textbackslash}apj, 683, 1104

\bibitem[{{Freedman} {et~al.}(2014){Freedman}, {Lustig-Yaeger}, {Fortney}, {Lupu}, {Marley}, \& {Lodders}}]{2014ApJS..214...25F}
{Freedman}, R.~S., {Lustig-Yaeger}, J., {Fortney}, J.~J., {et~al.} 2014, \apjs, 214, 25

\bibitem[{{Freedman} {et~al.}(2008){Freedman}, {Marley}, \& {Lodders}}]{2008ApJS..174..504F}
{Freedman}, R.~S., {Marley}, M.~S., \& {Lodders}, K. 2008, \apjs, 174, 504

\bibitem[{{Gabriel} {et~al.}(2014){Gabriel}, {Noels}, {Montalb{\'a}n}, \& {Miglio}}]{Gabriel2014}
{Gabriel}, M., {Noels}, A., {Montalb{\'a}n}, J., \& {Miglio}, A. 2014, \aap, 569, A63

\bibitem[{{Guillot}(2010)}]{2010A&A...520A..27G}
{Guillot}, T. 2010, \aap, 520, A27

\bibitem[{{Guillot} {et~al.}(1994){Guillot}, {Chabrier}, {Morel}, \& {Gautier}}]{1994Icar..112..354G}
{Guillot}, T., {Chabrier}, G., {Morel}, P., \& {Gautier}, D. 1994, \icarus, 112, 354

\bibitem[{Guillot \& Morel(1995)}]{guillot_cepam_1995}
Guillot, T. \& Morel, P. 1995, åps, 109, 109

\bibitem[{{Haldemann} {et~al.}(2020){Haldemann}, {Alibert}, {Mordasini}, \& {Benz}}]{2020A&A...643A.105H}
{Haldemann}, J., {Alibert}, Y., {Mordasini}, C., \& {Benz}, W. 2020, \aap, 643, A105

\bibitem[{{Hansen} \& {Kawaler}(1994)}]{1994sipp.book.....H}
{Hansen}, C.~J. \& {Kawaler}, S.~D. 1994, {Stellar Interiors. Physical Principles, Structure, and Evolution.}

\bibitem[{Harris {et~al.}(2020)Harris, Millman, van~der Walt, Gommers, Virtanen, Cournapeau, Wieser, Taylor, Berg, Smith, Kern, Picus, Hoyer, van Kerkwijk, Brett, Haldane, del R{\'{i}}o, Wiebe, Peterson, G{\'{e}}rard-Marchant, Sheppard, Reddy, Weckesser, Abbasi, Gohlke, \& Oliphant}]{harris2020array}
Harris, C.~R., Millman, K.~J., van~der Walt, S.~J., {et~al.} 2020, Nature, 585, 357

\bibitem[{{Helled} {et~al.}(2014){Helled}, {Bodenheimer}, {Podolak}, {Boley}, {Meru}, {Nayakshin}, {Fortney}, {Mayer}, {Alibert}, \& {Boss}}]{2014prpl.conf..643H}
{Helled}, R., {Bodenheimer}, P., {Podolak}, M., {et~al.} 2014, in Protostars and Planets VI, ed. H.~{Beuther}, R.~S. {Klessen}, C.~P. {Dullemond}, \& T.~{Henning}, 643--665

\bibitem[{{Helled} \& {Howard}(2024)}]{2024arXiv240705853H}
{Helled}, R. \& {Howard}, S. 2024, arXiv e-prints, arXiv:2407.05853

\bibitem[{{Heng} {et~al.}(2012){Heng}, {Hayek}, {Pont}, \& {Sing}}]{2012MNRAS.420...20H}
{Heng}, K., {Hayek}, W., {Pont}, F., \& {Sing}, D.~K. 2012, \mnras, 420, 20

\bibitem[{{Howard} \& {Guillot}(2023)}]{2023A&A...672L...1H}
{Howard}, S. \& {Guillot}, T. 2023, \aap, 672, L1

\bibitem[{Howard {et~al.}(2023)Howard, Guillot, Markham, Helled, Müller, Stevenson, Lunine, Miguel, \& Nettelmann}]{howard_exploring_2023}
Howard, S., Guillot, T., Markham, S., {et~al.} 2023, åp, 680, L2

\bibitem[{{Howard} {et~al.}(2025){Howard}, {Helled}, \& {M{\"u}ller}}]{howard_giant_2024}
{Howard}, S., {Helled}, R., \& {M{\"u}ller}, S. 2025, \aap, 693, L7

\bibitem[{{Howard} {et~al.}(2024){Howard}, {M{\"u}ller}, \& {Helled}}]{Howard2024}
{Howard}, S., {M{\"u}ller}, S., \& {Helled}, R. 2024, \aap, 689, A15

\bibitem[{{Hubbard} {et~al.}(2002){Hubbard}, {Burrows}, \& {Lunine}}]{2002ARA&A..40..103H}
{Hubbard}, W.~B., {Burrows}, A., \& {Lunine}, J.~I. 2002, \araa, 40, 103

\bibitem[{Hunter(2007)}]{Hunter2007}
Hunter, J.~D. 2007, Computing in Science \& Engineering, 9, 90

\bibitem[{{Iglesias} \& {Rogers}(1996)}]{1996ApJ...464..943I}
{Iglesias}, C.~A. \& {Rogers}, F.~J. 1996, \apj, 464, 943

\bibitem[{{Jermyn} {et~al.}(2023){Jermyn}, {Bauer}, {Schwab}, {Farmer}, {Ball}, {Bellinger}, {Dotter}, {Joyce}, {Marchant}, {Mombarg}, {Wolf}, {Sunny Wong}, {Cinquegrana}, {Farrell}, {Smolec}, {Thoul}, {Cantiello}, {Herwig}, {Toloza}, {Bildsten}, {Townsend}, \& {Timmes}}]{2023ApJS..265...15J}
{Jermyn}, A.~S., {Bauer}, E.~B., {Schwab}, J., {et~al.} 2023, \apjs, 265, 15

\bibitem[{{Joblib Development Team}(2020)}]{joblib}
{Joblib Development Team}. 2020, Joblib: running Python functions as pipeline jobs

\bibitem[{{Kippenhahn} {et~al.}(2013){Kippenhahn}, {Weigert}, \& {Weiss}}]{2013sse..book.....K}
{Kippenhahn}, R., {Weigert}, A., \& {Weiss}, A. 2013, {Stellar Structure and Evolution} ({Springer Berlin, Heidelberg})

\bibitem[{Kluyver {et~al.}(2016)Kluyver, Ragan-Kelley, P{\'e}rez, Granger, Bussonnier, Frederic, Kelley, Hamrick, Grout, Corlay, Ivanov, Avila, Abdalla, Willing, \& development team}]{jupyter}
Kluyver, T., Ragan-Kelley, B., P{\'e}rez, F., {et~al.} 2016, in Positioning and Power in Academic Publishing: Players, Agents and Agendas, ed. F.~Loizides \& B.~Scmidt (Netherlands: IOS Press), 87--90

\bibitem[{{Knierim}(2024{\natexlab{a}})}]{mesaCustomEoS}
{Knierim}, H. 2024{\natexlab{a}}, {Henrik-Knierim/mesa\_custom\_EoS: Version 1.0.0}

\bibitem[{{Knierim}(2024{\natexlab{b}})}]{pyMesaHelper}
{Knierim}, H. 2024{\natexlab{b}}, {Henrik-Knierim/py\_mesa\_helper: Version 1.0.0}

\bibitem[{{Knierim} \& {Helled}(2024)}]{Knierim2024}
{Knierim}, H. \& {Helled}, R. 2024, \apj, 977, 227

\bibitem[{{Knierim} \& {Helled}(2025)}]{Knierim2025}
{Knierim}, H. \& {Helled}, R. 2025, \aap, 698, L1

\bibitem[{{Kovetz} {et~al.}(2009){Kovetz}, {Yaron}, \& {Prialnik}}]{Kovetz2009}
{Kovetz}, A., {Yaron}, O., \& {Prialnik}, D. 2009, \mnras, 395, 1857

\bibitem[{{Liu} {et~al.}(2019){Liu}, {Hori}, {M{\"u}ller}, {Zheng}, {Helled}, {Lin}, \& {Isella}}]{2019Natur.572..355L}
{Liu}, S.-F., {Hori}, Y., {M{\"u}ller}, S., {et~al.} 2019, \nat, 572, 355

\bibitem[{Lorenzen {et~al.}(2011)Lorenzen, Holst, \& Redmer}]{Lorenzen2011}
Lorenzen, W., Holst, B., \& Redmer, R. 2011, Phys. Rev. B, 84, 235109

\bibitem[{Lyon(1978)}]{lyon1978sesame}
Lyon, S.~P. 1978, LANL report

\bibitem[{{Marigo} {et~al.}(2024){Marigo}, {Addari}, {Bossini}, {Bressan}, {Costa}, {Girardi}, {Pastorelli}, {Trabucchi}, \& {Volpato}}]{2024ApJ...976...39M}
{Marigo}, P., {Addari}, F., {Bossini}, D., {et~al.} 2024, \apj, 976, 39

\bibitem[{{Marigo} \& {Aringer}(2009)}]{2009A&A...508.1539M}
{Marigo}, P. \& {Aringer}, B. 2009, \aap, 508, 1539

\bibitem[{{Marigo} {et~al.}(2022){Marigo}, {Aringer}, {Girardi}, \& {Bressan}}]{2022ApJ...940..129M}
{Marigo}, P., {Aringer}, B., {Girardi}, L., \& {Bressan}, A. 2022, \apj, 940, 129

\bibitem[{Marley {et~al.}(2007)Marley, Fortney, Hubickyj, Bodenheimer, \& Lissauer}]{marley_luminosity_2007}
Marley, M.~S., Fortney, J.~J., Hubickyj, O., Bodenheimer, P., \& Lissauer, J.~J. 2007, {\textbackslash}apj, 655, 541

\bibitem[{{Meier} {et~al.}(2025){Meier}, {Reinhardt}, {Shibata}, {M{\"u}ller}, {Stadel}, \& {Helled}}]{2025arXiv250323997M}
{Meier}, T., {Reinhardt}, C., {Shibata}, S., {et~al.} 2025, \apj, 988, 7

\bibitem[{{Mol Lous} {et~al.}(2024){Mol Lous}, {Mordasini}, \& {Helled}}]{2024A&A...685A..22M}
{Mol Lous}, M., {Mordasini}, C., \& {Helled}, R. 2024, \aap, 685, A22

\bibitem[{{More} {et~al.}(1988){More}, {Warren}, {Young}, \& {Zimmerman}}]{1988PhFl...31.3059M}
{More}, R.~M., {Warren}, K.~H., {Young}, D.~A., \& {Zimmerman}, G.~B. 1988, Physics of Fluids, 31, 3059

\bibitem[{Movshovitz {et~al.}(2010)Movshovitz, Bodenheimer, Podolak, \& Lissauer}]{movshovitz_formation_2010}
Movshovitz, N., Bodenheimer, P., Podolak, M., \& Lissauer, J.~J. 2010, ıcarus, 209, 616

\bibitem[{M{\"u}ller(2021)}]{zora230638}
M{\"u}ller, S. 2021, PhD thesis, University of Zurich

\bibitem[{{M{\"u}ller} {et~al.}(2020{\natexlab{a}}){M{\"u}ller}, {Ben-Yami}, \& {Helled}}]{2020ApJ...903..147M}
{M{\"u}ller}, S., {Ben-Yami}, M., \& {Helled}, R. 2020{\natexlab{a}}, \apj, 903, 147

\bibitem[{{M{\"u}ller} \& {Helled}(2024)}]{2024ApJ...967....7M}
{M{\"u}ller}, S. \& {Helled}, R. 2024, \apj, 967, 7

\bibitem[{{M{\"u}ller} \& {Helled}(2025)}]{2025A&A...693L...4M}
{M{\"u}ller}, S. \& {Helled}, R. 2025, \aap, 693, L4

\bibitem[{{M{\"u}ller} {et~al.}(2020{\natexlab{b}}){M{\"u}ller}, {Helled}, \& {Cumming}}]{2020A&A...638A.121M}
{M{\"u}ller}, S., {Helled}, R., \& {Cumming}, A. 2020{\natexlab{b}}, \aap, 638, A121

\bibitem[{Müller(2024)}]{pymesatools}
Müller, S. 2024, mesatools, \url{https://github.com/tiny-hippo/pymesatools}

\bibitem[{Müller(2025)}]{tinyeos}
Müller, S. 2025, tinyeos, \url{https://github.com/tiny-hippo/tinyeos}

\bibitem[{Müller \& Helled(2021)}]{muller_synthetic_2021}
Müller, S. \& Helled, R. 2021, {\textbackslash}mnras, 507, 2094

\bibitem[{Müller \& Helled(2023)}]{muller_towards_2023}
Müller, S. \& Helled, R. 2023, åp, 669, A24

\bibitem[{{Parmentier} {et~al.}(2015){Parmentier}, {Guillot}, {Fortney}, \& {Marley}}]{2015A&A...574A..35P}
{Parmentier}, V., {Guillot}, T., {Fortney}, J.~J., \& {Marley}, M.~S. 2015, \aap, 574, A35

\bibitem[{{Paxton} {et~al.}(2011){Paxton}, {Bildsten}, {Dotter}, {Herwig}, {Lesaffre}, \& {Timmes}}]{2011ApJS..192....3P}
{Paxton}, B., {Bildsten}, L., {Dotter}, A., {et~al.} 2011, \apjs, 192, 3

\bibitem[{{Paxton} {et~al.}(2013){Paxton}, {Cantiello}, {Arras}, {Bildsten}, {Brown}, {Dotter}, {Mankovich}, {Montgomery}, {Stello}, {Timmes}, \& {Townsend}}]{2013ApJS..208....4P}
{Paxton}, B., {Cantiello}, M., {Arras}, P., {et~al.} 2013, \apjs, 208, 4

\bibitem[{{Paxton} {et~al.}(2015){Paxton}, {Marchant}, {Schwab}, {Bauer}, {Bildsten}, {Cantiello}, {Dessart}, {Farmer}, {Hu}, {Langer}, {Townsend}, {Townsley}, \& {Timmes}}]{2015ApJS..220...15P}
{Paxton}, B., {Marchant}, P., {Schwab}, J., {et~al.} 2015, \apjs, 220, 15

\bibitem[{{Paxton} {et~al.}(2019){Paxton}, {Smolec}, {Schwab}, {Gautschy}, {Bildsten}, {Cantiello}, {Dotter}, {Farmer}, {Goldberg}, {Jermyn}, {Kanbur}, {Marchant}, {Thoul}, {Townsend}, {Wolf}, {Zhang}, \& {Timmes}}]{2019ApJS..243...10P}
{Paxton}, B., {Smolec}, R., {Schwab}, J., {et~al.} 2019, \apjs, 243, 10

\bibitem[{{Podolak} {et~al.}(2023){Podolak}, {Levi}, {Vazan}, \& {Malamud}}]{2023Icar..39415424P}
{Podolak}, M., {Levi}, A., {Vazan}, A., \& {Malamud}, U. 2023, \icarus, 394, 115424

\bibitem[{{Poser} {et~al.}(2019){Poser}, {Nettelmann}, \& {Redmer}}]{2019Atmos..10..664P}
{Poser}, A.~J., {Nettelmann}, N., \& {Redmer}, R. 2019, Atmosphere, 10, 664

\bibitem[{{Poser} \& {Redmer}(2024)}]{2024MNRAS.529.2242P}
{Poser}, A.~J. \& {Redmer}, R. 2024, \mnras, 529, 2242

\bibitem[{{Rauer} {et~al.}(2025){Rauer}, {Aerts}, {Cabrera}, {Deleuil}, {Erikson}, {Gizon}, {Goupil}, {Heras}, {Walloschek}, {Lorenzo-Alvarez}, {Marliani}, {Martin-Garcia}, {Mas-Hesse}, {O'Rourke}, {Osborn}, {Pagano}, {Piotto}, {Pollacco}, {Ragazzoni}, {Ramsay}, {Udry}, {Appourchaux}, {Benz}, {Brandeker}, {G{\"u}del}, {Janot-Pacheco}, {Kabath}, {Kjeldsen}, {Min}, {Santos}, {Smith}, {Suarez}, {Werner}, {Aboudan}, {Abreu}, {Acu{\~n}a}, {Adams}, {Adibekyan}, {Affer}, {Agneray}, {Agnor}, {Aguirre B{\o}rsen-Koch}, {Ahmed}, {Aigrain}, {Al-Bahlawan}, {Alcacera Gil}, {Alei}, {Alencar}, {Alexander}, {Alfonso-Garz{\'o}n}, {Alibert}, {Allende Prieto}, {Almeida}, {Alonso Sobrino}, {Altavilla}, {Althaus}, {Alvarez Trujillo}, {Amarsi}, {Ammler-von Eiff}, {Am{\^o}res}, {Andrade}, {Antoniadis-Karnavas}, {Ant{\'o}nio}, {Aparicio del Moral}, {Appolloni}, {Arena}, {Armstrong}, {Aroca Aliaga}, {Asplund}, {Audenaert}, {Auricchio}, {Avelino}, {Baeke}, {Bailli{\'e}}, {Balado}, {Ballber Balaguer{\'o}}, {Balestra}, {Ball}, {Ballans},
  {Ballot}, {Barban}, {Barbary}, {Barbieri}, {Barcel{\'o} Forteza}, {Barker}, {Barklem}, {Barnes}, {Barrado Navascues}, {Barragan}, {Baruteau}, {Basu}, {Baudin}, {Baumeister}, {Bayliss}, {Bazot}, {Beck}, {Belkacem}, {Bellinger}, {Benatti}, {Benomar}, {B{\'e}rard}, {Bergemann}, {Bergomi}, {Bernardo}, {Biazzo}, {Bignamini}, {Bigot}, {Billot}, {Binet}, {Biondi}, {Biondi}, {Birch}, {Bitsch}, {Bluhm Ceballos}, {B{\'o}di}, {Bogn{\'a}r}, {Boisse}, {Bolmont}, {Bonanno}, {Bonavita}, {Bonfanti}, {Bonfils}, {Bonito}, {Bonomo}, {B{\"o}rner}, {Boro Saikia}, {Borreguero Mart{\'\i}n}, {Borsa}, {Borsato}, {Bossini}, {Bouchy}, {Bou{\'e}}, {Boufleur}, {Boumier}, {Bourrier}, {Bowman}, {Bozzo}, {Bradley}, {Bray}, {Bressan}, {Breton}, {Brienza}, {Brito}, {Brogi}, {Brown}, {Brown}, {Brun}, {Bruno}, {Bruns}, {Buchhave}, {Bugnet}, {Buldgen}, {Burgess}, {Busatta}, {Busso}, {Buzasi}, {Caballero}, {Cabral}, {Cabrero Gomez}, {Calderone}, {Cameron}, {Cameron}, {Campante}, {Campos Gestal}, {Canto Martins}, {Cara}, {Carone}, {Carrasco},
  {Casagrande}, {Casewell}, {Cassisi}, {Castellani}, {Castro}, {Catala}, {Catal{\'a}n Fern{\'a}ndez}, {Catelan}, {Cegla}, {Cerruti}, {Cessa}, {Chadid}, {Chaplin}, {Charpinet}, {Chiappini}, {Chiarucci}, {Chiavassa}, {Chinellato}, {Chirulli}, {Christensen-Dalsgaard}, {Church}, {Claret}, {Clarke}, {Claudi}, {Clermont}, {Coelho}, {Coelho}, {Cogato}, {Colom{\'e}}, {Condamin}, {Conde Garc{\'\i}a}, \& {Conseil}}]{2025ExA....59...26R}
{Rauer}, H., {Aerts}, C., {Cabrera}, J., {et~al.} 2025, Experimental Astronomy, 59, 26

\bibitem[{{Saumon} {et~al.}(1995){Saumon}, {Chabrier}, \& {van Horn}}]{1995ApJS...99..713S}
{Saumon}, D., {Chabrier}, G., \& {van Horn}, H.~M. 1995, \apjs, 99, 713

\bibitem[{{Sch{\"o}ttler} \& {Redmer}(2018)}]{Schoettler2018}
{Sch{\"o}ttler}, M. \& {Redmer}, R. 2018, \prl, 120, 115703

\bibitem[{{Shibata} \& {Helled}(2025)}]{2025arXiv250701212S}
{Shibata}, S. \& {Helled}, R. 2025, \aap, 700, A224

\bibitem[{{Siebenaler} {et~al.}(2025){Siebenaler}, {Miguel}, {de Regt}, \& {Guillot}}]{2025A&A...693A.308S}
{Siebenaler}, L., {Miguel}, Y., {de Regt}, S., \& {Guillot}, T. 2025, \aap, 693, A308

\bibitem[{{Stevenson} \& {Salpeter}(1977)}]{Stevenson_1977b}
{Stevenson}, D.~J. \& {Salpeter}, E.~E. 1977, \apjs, 35, 239

\bibitem[{{Sur} {et~al.}(2024){Sur}, {Su}, {Tejada Arevalo}, {Chen}, \& {Burrows}}]{2024ApJ...971..104S}
{Sur}, A., {Su}, Y., {Tejada Arevalo}, R., {Chen}, Y.-X., \& {Burrows}, A. 2024, \apj, 971, 104

\bibitem[{{Sur} {et~al.}(2025){Sur}, {Tejada Arevalo}, {Su}, \& {Burrows}}]{Sur2025}
{Sur}, A., {Tejada Arevalo}, R., {Su}, Y., \& {Burrows}, A. 2025, \apjl, 980, L5

\bibitem[{Teske {et~al.}(2019)Teske, Thorngren, Fortney, Hinkel, \& Brewer}]{teske_metal-rich_2019}
Teske, J.~K., Thorngren, D., Fortney, J.~J., Hinkel, N., \& Brewer, J.~M. 2019, {\textbackslash}aj, 158, 239

\bibitem[{Thorngren {et~al.}(2016)Thorngren, Fortney, Murray-Clay, \& Lopez}]{thorngren_mass-metallicity_2016}
Thorngren, D.~P., Fortney, J.~J., Murray-Clay, R.~A., \& Lopez, E.~D. 2016, {\textbackslash}apj, 831, 64

\bibitem[{{Valencia} {et~al.}(2013){Valencia}, {Guillot}, {Parmentier}, \& {Freedman}}]{2013ApJ...775...10V}
{Valencia}, D., {Guillot}, T., {Parmentier}, V., \& {Freedman}, R.~S. 2013, \apj, 775, 10

\bibitem[{Valletta \& Helled(2020)}]{valletta_giant_2020}
Valletta, C. \& Helled, R. 2020, {\textbackslash}apj, 900, 133

\bibitem[{{Vazan} {et~al.}(2018){Vazan}, {Helled}, \& {Guillot}}]{Vazan2018}
{Vazan}, A., {Helled}, R., \& {Guillot}, T. 2018, \aap, 610, L14

\bibitem[{Vazan {et~al.}(2013)Vazan, Kovetz, Podolak, \& Helled}]{vazan_effect_2013}
Vazan, A., Kovetz, A., Podolak, M., \& Helled, R. 2013, {\textbackslash}mnras, 434, 3283

\bibitem[{Virtanen {et~al.}(2020)Virtanen, Gommers, Oliphant, Haberland, Reddy, Cournapeau, Burovski, Peterson, Weckesser, Bright, {van der Walt}, Brett, Wilson, Millman, Mayorov, Nelson, Jones, Kern, Larson, Carey, Polat, Feng, Moore, {VanderPlas}, Laxalde, Perktold, Cimrman, Henriksen, Quintero, Harris, Archibald, Ribeiro, Pedregosa, {van Mulbregt}, \& {SciPy 1.0 Contributors}}]{2020SciPy-NMeth}
Virtanen, P., Gommers, R., Oliphant, T.~E., {et~al.} 2020, Nature Methods, 17, 261

\bibitem[{{Visscher} {et~al.}(2006){Visscher}, {Lodders}, \& {Fegley}}]{2006ApJ...648.1181V}
{Visscher}, C., {Lodders}, K., \& {Fegley}, Jr., B. 2006, \apj, 648, 1181

\bibitem[{{Visscher} {et~al.}(2010){Visscher}, {Lodders}, \& {Fegley}}]{2010ApJ...716.1060V}
{Visscher}, C., {Lodders}, K., \& {Fegley}, Jr., B. 2010, \apj, 716, 1060

\bibitem[{{W}es {M}c{K}inney(2010)}]{mckinney-proc-scipy-2010}
{W}es {M}c{K}inney. 2010, in {P}roceedings of the 9th {P}ython in {S}cience {C}onference, ed. {S}t\'efan van~der {W}alt \& {J}arrod {M}illman, 56 -- 61

\bibitem[{Wolf(2025)}]{py_mesa_reader}
Wolf, B. 2025, mesa\_reader, \url{https://github.com/wmwolf/py_mesa_reader}

\end{thebibliography}

\appendix

\section{Equation of state details}\label{sec:appendix_equation_of_state}

In Table \ref{tab:eos_quantities} we list the EoS quantities required by \mesa. In Section \ref{sec:mesa_eos}, we explained how the density, pressure, entropy and internal energy are determined. Here, we also describe how the other quantities are calculated when using the recommended density-temperature EoS, which employs the pressure-temperature tables and performs a root-finding procedure. In the following description, the quantities of the individual components are marked by the index $i$, and the partial derivatives involved are calculated directly from the bicubic splines -- no finite differences are used.

With the aid of the inverse-function and triple-product rules, the pressure derivatives are calculated as  

\begin{subequations}
    \begin{align}
        \chi_\rho &= \left(\altpder{\ln \rho}{\ln p}{T} \, \, \right)^{-1} \, , \\
        % \notag \\
        \chi_T &= -\frac{\altpder{\ln \rho}{\ln T}{p}}{\altpder{\ln \rho}{\ln p}{T}} \, .
    \end{align}
\end{subequations}

\noindent
The partial derivatives of the density with respect to pressure and temperature are 

\begin{subequations}
    \begin{align}
        \pder{\ln \rho}{\ln p}{T} &= \rho \sum_i \frac{X_i }{\rho_i \,\chi_{\rho, i}} \, , \\
         % \notag \\
        \pder{\ln \rho}{\ln T}{p} &= - \rho \sum_i \frac{X_i \, \chi_{T, i}}{\rho_i \, \chi_{\rho, i}} \, .
    \end{align}
\end{subequations}

\noindent
The adiabatic temperature gradient for the mixture is calculated using the triple-product rule as

\begin{equation}
    \nabla_{\rm{ad}} = - \frac{\altpder{\ln s}{\ln p}{T}}{\altpder{\ln s}{\ln T}{p}} \, ,
\end{equation}

\noindent
where the partial derivatives of the entropy with respect to pressure and temperature are  

\begin{subequations}
    \begin{align}
            \pder{\ln s}{\ln p}{T} &= s^{-1} \sum_i X_i s_i \pder{\ln s_i}{\ln p}{T} \, , \\
            % \notag \\
            \pder{\ln s}{\ln T}{p} &= s^{-1} \sum_i X_i s_i \pder{\ln s}{\ln T}{p} \, .
    \end{align}
\end{subequations}

\noindent
The other required quantities related to derivatives are calculated using the two pressure derivatives and the adiabatic temperature gradient (by employing the triple-product rule and Maxwell relations; see, for example, \citet{1994sipp.book.....H}). The specific heat capacities at constant pressure and volume are  

\begin{subequations}
    \begin{align}
        c_p &= \frac{p \, \chi_T}{\rho T \, \chi_\rho \nabla_{\rm{ad}}} \, , \\
        % \notag \\
        c_V &= \frac{c_p \, \chi_\rho}{\Gamma_1} \, .
    \end{align}
\end{subequations}

\noindent
The adiabatic indices $\Gamma_1$ and $\Gamma_3$ are calculated as

\begin{subequations}
    \begin{align}
        \Gamma_1 &= \frac{\chi_\rho}{1 - \chi_T \, \nabla_{\rm{ad}}} \, , \\
        % \notag \\
        \Gamma_3 &= 1 + \Gamma_1 \nabla_{\rm{ad}} \, .
    \end{align}
\end{subequations}

\noindent
The three remaining partial derivatives are

\begin{subequations}
    \begin{align}
        \pder{s}{\rho}{T} &= \frac{c_V}{T} \, , \\
        % \notag \\
        \pder{s}{T}{\rho} &= - \frac{p \, \chi_T}{T \rho^2} \, , \\
        % \notag \\
        \pder{u}{\rho}{T} &= (1 - \chi_T) \frac{ p}{\rho^2} \, .
    \end{align}
\end{subequations}

\noindent
While the following three quantities are generally not relevant for standard giant planet evolution models, for completeness, we attempt to approximate them as well as possible given the current tabulated equations of state.

Calculating the mean molecular weight of the mixture requires knowledge of the number fractions of the molecules and ions involved in the mixture. Unfortunately, neither the \citet{2021ApJ...917....4C} hydrogen-helium EoS nor the heavy-element tables include this information. However, these are tabulated in the \citet{1995ApJS...99..713S} EoS, which we use to calculate the mean molecular weight of hydrogen ($\mu_X$), helium ($\mu_Y$) and their mixture ($\mu_{X, Y}$) 

\begin{subequations}
    \begin{align}
        \frac{1}{\mu_{\rm{H}}} &= \frac{2 X}{1 + x_{\rm{H}} + + 3 x_{\rm{H_2}}} \, , \\
        \frac{1}{\mu_{\rm{He}}}  &= \frac{3 Y}{4 (1 + 2x_{\rm{He}} + + 3 x_{\rm{He^+}})} \, , \\
        \frac{1}{\mu_{\rm{H, He}}}  &= \frac{1}{\mu_{\rm{H}}} + \frac{1}{\mu_{\rm{He}}} \, ,
    \end{align}
\end{subequations}

\noindent
where the $x_i$ are the number fractions of the atoms, molecules or ions. For the heavy elements, we approximate their mean molecular weight using the atomic masses $m_a$ of the involved chemical species, such that the total mean-molecular weight is, 

\begin{equation}
    \frac{1}{\mu} = \frac{Z}{m_a} + \frac{1 - Z}{\mu_{\rm{H, He}}} \, .
\end{equation}

\noindent
Similarly, ignoring the heavy-element contribution and again using the \citet{1995ApJS...99..713S} hydrogen-helium EoS, the mean number of free electrons per nucleon is,  

\begin{equation}
    \frac{1}{\mu_{\rm{e}}} = \frac{x_{\rm{e, H}}}{\mu_{\rm{H}}} + \frac{x_{\rm{e, He}}}{\mu_{\rm{He}}} \, ,
\end{equation}

\noindent
where the $x_{\rm{e, i}}$ are the number fractions of free electrons contributed by hydrogen or helium.

Finally, the inverse chemical potential of the free electrons $\eta$ (degeneracy parameter) is calculated from the mean number of free electrons per nucleon and with the rational function approximation of the inverse Fermi-Dirac integral from \citet{1993ApJS...84..101A}. 

\begin{table}[]
    \centering
    \caption{EoS in- and outputs for the \mesa tables, their definitions and units; also see \citet{2011ApJS..192....3P} and \citep{zora230638}.}
    \begin{tabular}{ccc}
        \toprule
        Input  & Definition & Units \\
        \midrule
        $T$ & Temperature & K \\
        $\rho$ & Density & g cm$^{-3}$ \\
        \midrule
        Output & & \\
        \midrule
        $p$ & Gas pressure  & erg cm$^{-3}$ \\
        $s$ & Entropy per gram & erg g$^{-1}$ K$^{-1}$ \\
        $u$ & Internal energy per gram & erg g$^{-1}$ \\
        $\chi_{\rho}$ & $\equiv \pder{\ln p}{\ln \rho}{T}$ & None \\
        $\chi_{T}$ & $\equiv \pder{\ln p}{\ln T}{\rho}$ & None \\
        $\nabla_{\rm{ad}}$ & $\equiv \pder{\ln T}{\ln p}{s}$ & None \\
        $c_{p}$ & \makecell{Specific heat \\ at constant pressure} & erg g$^{-1}$ K$^{-1}$ \\
        $c_{V}$ & \makecell{Specific heat \\ at constant $V \equiv 1/\rho$} & erg g$^{-1}$ K$^{-1}$ \\
        $\Gamma_{1}$ & $\equiv \pder{\ln p}{\ln \rho}{s}$ & None \\
        $\Gamma_{3}$ & $\equiv \pder{\ln T}{\ln \rho}{s} + 1$ & None \\
        $\pder{s}{\rho}{T}$ & \textemdash & erg cm$^{3}$ g$^{-2}$ K$^{-1}$ \\
        $\pder{s}{T}{\rho}$ & \textemdash & erg g$^{-1}$ K$^{-2}$ \\
        $\pder{u}{\rho}{T}$ & \textemdash & erg cm$^{3}$ g$^{-2}$ \\
        $\mu$ & \makecell{Mean molecular weight \\ per gas particle} & None \\
        $\eta$ & \makecell{Ratio of electron \\ chemical potential to $k_{b} T$} & None \\
        $1/\mu_{e}$ & \makecell{Mean number of \\ free electrons per nucleon} & None \\
        \bottomrule
    \end{tabular}
    \label{tab:eos_quantities}
\end{table}

%
% \newpage
\section{Effect of mixing on the inferred temperature profile}\label{sec:appendix_gentle_mixing}
The assumed internal structure and whether convective mixing is included significantly affect the temperature profile within the planet. 
Figure \ref{fig:mixing_temperature_comparison} compares the resulting temperature profiles for simulations with and without mixing after \SI{10}{\Gyr}. 
\begin{figure}
  \centering
  \includegraphics[width=0.9\columnwidth]{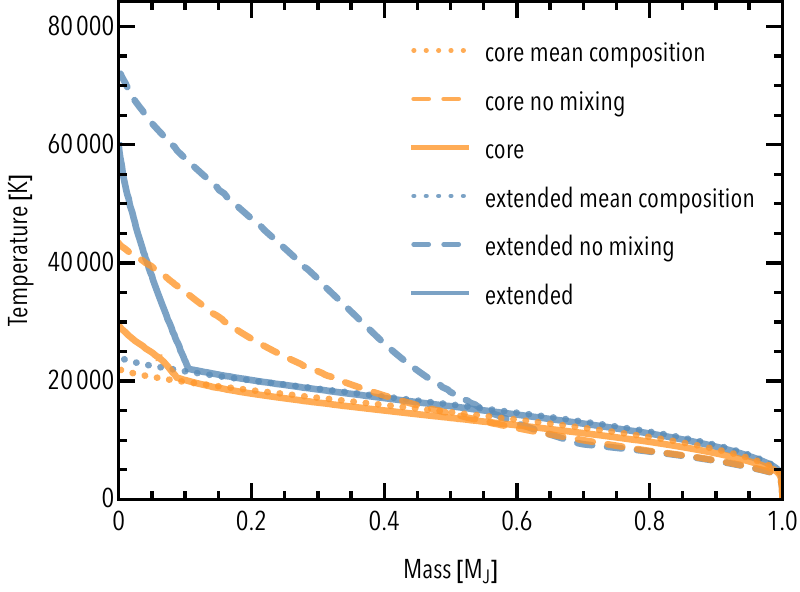}
  \caption{Comparison of the inferred temperature profiles after \SI{10}{\Gyr} for models with mixing, without mixing, and for a homogeneous profile with the same bulk metallicity as the corresponding composition gradient. All the models were run under the same assumptions as in Fig. \ref{fig:Z_evolution}.}
  \label{fig:mixing_temperature_comparison}
\end{figure}
If mixing is ignored, composition gradients artificially suppress convection, resulting in interiors that are considerably hotter than they should be.

\section{Atmospheric helium mass fraction with helium rain }\label{sec:helium_fraction}
 
As helium rains and helium droplets settle down to deeper layers, the planetary bulk composition has to be conserved. However, there is more than one way to ensure constant composition and it is unclear which treatment is most appropriate.

For example, let's assume that the primordial composition of the atmosphere is $X = 0.7$, $Y = 0.28$, $Z = 0.02$, and at present, $Y = 0.1$. To keep $X + Y + Z = 1$, one can choose between the following options: 

\begin{enumerate}
    \item Assume that $(X / Z)_{\rm{primordial}}=(X/Z)_{\rm{new}}$ and with having $Y_{\rm{new}} = 0.1$, one can find $X_{\rm{new}}$ and $Z_{\rm{new}}$.
    \item If $Y_{\rm{new}} =0.1$, we can assume that 18\% of helium that rained down is replaced by hydrogen, so that $X_{\rm{new}} = 0.88$ and $Z_{\rm{primordial}} = Z_{\rm{new}} = 0.02$. In this case, however, the local hydrogen-to-metals ratio changes.
    \item Assume that the missing 18\% are of solar composition. In this case, however, ``new" helium is brought to the atmosphere. 
\end{enumerate}

In our simulations with the \texttt{instant\_rain} algorithm, we used the first option corresponding to a rain of pure helium. This preserves the ratio of hydrogen to heavy elements since it is the same in both the material from which the rain is formed and the material replacing the droplets (by convection from below). In that case, if we start with  $X = 0.7$, $Y = 0.28$, and $Z = 0.02$ the atmospheric composition at present-day would be $X = 0.875$,
$Y = 0.1$, and $Z = 0.025$.

We note that each option would lead to different results. In addition, reality is surely more complicated: Raindrops could also take down some hydrogen and heavy elements \cite[for example, neon][]{Stevenson_1977b}. In addition, a slight enrichment of heavy elements in the envelope may be expected as most of the species of interest (such as water) prefer metallic hydrogen over helium. 

We suggest that future research should be dedicated to modeling the planetary evolution accounting for elemental mixing and phase separation self-consistently. This would require conserving the total mass of each species individually while ensuring that at every timestep, the phase diagram (which has multiple dimensions) is satisfied, determining  $X$, $Y$ and $Z$ of the material being left behind as
raindrops form and material is convected up from below. 

\end{document}